\documentclass[ejs]{imsart}

\RequirePackage[OT1]{fontenc}
\RequirePackage{amsthm,amsmath}
\RequirePackage{natbib}
\RequirePackage[colorlinks,citecolor=blue,urlcolor=blue]{hyperref}

\usepackage{amsthm,amsmath}
\usepackage{amssymb}
\usepackage{amsfonts}
\usepackage[english,ruled,vlined]{algorithm2e}
\usepackage{graphicx}
\usepackage{hypernat}
\usepackage{subfig}
\usepackage{multirow}

\pubyear{2006}
\volume{0}
\issue{0}
\firstpage{1}
\lastpage{8}

\newcommand{\cst}{\mathrm{const}}

\newcommand{\Esp}{\mathrm{E}}

\newcommand{\Bernoulli}{\mathcal{B}}
\newcommand{\Gaussian}{\mathcal{N}}

\newcommand{\KL}{\mathrm{KL}}
\newcommand{\LowerBound}{{\mathcal L}}
\newcommand{\Trace}{\mathrm{Tr}}

\newcommand{\asvec}{\mathrm{vec}}

\newcommand{\ttau}{ \tilde{\tau }}

\DeclareMathOperator{\covmat}{{\bf S}}
\DeclareMathOperator{\bZ}{{\bf Z}}
\DeclareMathOperator{\bz}{{\bf z}}
\DeclareMathOperator{\tZ}{\tilde{Z}}
\DeclareMathOperator{\btZ}{\tilde{\bZ}}
\DeclareMathOperator{\bI}{{\bf I}}

\DeclareMathOperator{\bX}{{\bf X}}

\DeclareMathOperator{\bA}{\bf A}

\DeclareMathOperator{\bB}{\bf B}
\DeclareMathOperator{\bW}{{\bf W}}
\DeclareMathOperator{\tW}{\tilde{W}}
\DeclareMathOperator{\btW}{\tilde{\bW}}

\DeclareMathOperator{\bU}{{\bf U}}

\DeclareMathOperator{\bV}{{\bf V}}

\DeclareMathOperator{\bE}{{\bf E}}
\DeclareMathOperator{\btE}{\tilde{\bE}}
\DeclareMathOperator{\Beta}{Beta}
\DeclareMathOperator{\Gam}{Gam}

\DeclareMathOperator{\btheta}{\boldsymbol{\theta}}
\DeclareMathOperator{\balpha}{\boldsymbol{\alpha}}

\DeclareMathOperator{\bSigma}{\boldsymbol{\Sigma}}
\DeclareMathOperator{\bxi}{\boldsymbol{\xi}}
\DeclareMathOperator{\bpi}{\boldsymbol{\pi}}
\DeclareMathOperator{\bPi}{\boldsymbol{\Pi}}

\DeclareMathOperator{\btau}{\boldsymbol{\tau}}
\DeclareMathOperator{\bttau}{\tilde{\btau}}

\startlocaldefs
\numberwithin{equation}{section}
\theoremstyle{plain}
\newtheorem{thm}{Theorem}[section]
\newtheorem{prop}{Proposition}[section]
\endlocaldefs

\graphicspath{{figures/}}

\begin{document}

\begin{frontmatter}
\title{Model Selection in Overlapping Stochastic Block Models}
\runtitle{Model Selection in Overlapping Stochastic Block Models}

\begin{aug}
\author{\fnms{Pierre} \snm{Latouche}\ead[label=e1]{pierre.latouche@univ-paris1.fr}}

\address{Laboratoire   SAMM,   EA    4543   \\Universit\'e   Paris   1
  Panth\'eon--Sorbonne \\
\printead{e1}}

\author{\fnms{Etienne} \snm{Birmel\'{e}}\ead[label=e2]{etienne.birmele@parisdescartes.fr}}

\address{
  Laboratoire MAP5 \\
  Universit\'{e} Paris Descartes and CNRS, Sorbonne Paris Cit\'{e} \\
  \printead{e2}}

\author{\fnms{Christophe}
  \snm{Ambroise}\ead[label=e3]{christophe.ambroise@genopole.cnrs.fr}}

\address{
  Laboratoire Statistique et G\'{e}nome \\
  UMR CNRS 8071, INRA 1152, UEVE \\
  \printead{e3}}

\runauthor{P. Latouche et al.}

\affiliation{Universit\'e   Paris   1
  Panth\'eon--Sorbonne}

\end{aug}

\begin{abstract}
Networks are a commonly used mathematical model to describe the rich set of interactions between objects
of interest.  Many clustering methods have been developed in order to 
partition such structures, among which several rely
on underlying probabilistic models, typically mixture models. The relevant hidden structure may however show overlapping groups in several applications.
The Overlapping Stochastic Block Model [Latouche, Birmel\'e and Ambroise (2011)] has been developed to take this phenomenon into account. Nevertheless, the  
problem of the  choice of the number of classes  in the inference step
is  still open.  To  tackle this  issue,  we  consider  the
proposed model  in a  Bayesian framework and  develop a  new criterion
based on a non asymptotic approximation of the marginal log-likelihood. We
describe how the criterion can be computed through a variational Bayes
EM algorithm, and demonstrate its efficiency by running it on both simulated and real data.
\end{abstract}


\tableofcontents
\end{frontmatter}

\section{Introduction}

Networks are commonly used to describe complex interaction patterns in different fields like social sciences \citep{articlesnijders1997} or biology \citep{articlealbert2002}. They provide a common  mathematical framework to study data sets as various as social relations \citep{articlepalla2007}, protein-protein interactions \citep{articlebarabasi2004} or the Internet \citep{articlezanghi2008}. One way to learn knowledge from such large data sets is to cluster their vertices according to their topological behaviour. Numerous probabilistic methods have been developed so far to achieve this goal according to different types of underlying models.

Most methods look for community structure, or assortative mixing, that
is  cluster the  vertices such  that vertices  of a  class  are mostly
connected        with        vertices        of        the        same
class. \cite{proceedingsgirvan2002}  propose to maximize  a modularity
score based  on the observed values  of the internal  densities of the
classes, compared  with their expected  values in a random  model. The
choice of the  optimal number of classes is  done by splitting current
classes   as   long   as   a   modularity   gain   can   be   achieved
\citep{proceedingsnewman2006}. However, algorithms based on modularity
are  asymptotically  biased  and   may  lead  to  incorrect  community
structures                 as                 shown                 by
\cite{proceedingsbickel2009}.  \cite{articlehandcock2007}  propose  to
map  the vertices in  a continuous  latent space  and to  cluster them
according to their positions. A maximum likelihood approach as well as
a Bayesian procedure, coupled with a BIC criterion to estimate the number of classes, are implemented in the R package latentnet \citep{manualkrivitsky2009}.  

The community structure assumption  is however not relevant in several
types  of networks.  Transcription  factors may  for example  regulate
common operons without regulating each other directly. Other examples
like   actors  or   citation   networks  even   exhibit  a   bipartite
structure.     \citet{articleestrada2005}    therefore     look    for
disassortative mixing, in which  most edges link vertices of different
classes.  \cite{articlehofman2008}  define   a  mixture  model  by  an
intra-group  connectivity $\lambda$  and  an inter-group  connectivity
$\epsilon$,   which  allows   to  deal   with  both   assortative  and
disassortative   mixing.   Moreover,   they  develop   a   variational
approximation of  the marginal  log-likelihood and use it to derive
a non asymptotic Bayesian criterion to estimate the number of classes. It is implemented in the software VBMOD.

The Stochastic Block Model (SBM) \citep{wang1987}, initially introduced in social
sciences   \citep{articlefienberg1981,articleholland1983},  allows  to
cluster  the   vertices  according  to  both   their  preferences  and
aversions. It assumes that the vertices of the network are spread into
$Q$ classes and that  the connection probabilities between the classes
are given by a $Q\times Q$ matrix $\bPi$ \citep{articlefrank1982}. Due
to the flexibility  of the connectivity pattern given  by $\bPi$, this
model  generalizes the  previous ones,  as  it can  deal with  network
structures which are  neither assortative nor disassortative. However,
the classical EM algorithm  \cite{articledempster1977} cannot be used
directly  as the posterior  distribution $p(\bZ  |\bX)$ of  the latent
class variables  $\bZ$ given the data  do not factorize.  To get round
this  difficulty, \cite{articlenowicki2001}  use  a Bayesian  approach
based   on   a   Gibbs    sampling   estimation   of   the   posterior
distributions.  This method  is  implemented in  the software  BLOCKS,
available in the package StoCNET \citep{manualboer2006}. However, no model
based   criterion    is   given    to   determine   the    number   of
classes.  \cite{articledaudin2008}  and  \cite{articlemariadassou2010}
propose to  tackle that  issue in a  frequentist framework  through an
asymptotic    approximation    of    the   integrated    complete-data
log-likelihood.   In a  Bayesian  framework, \cite{inbooklatouche2009}
introduce a non asymptotic approximation of the marginal log-likelihood as a criterion to estimate the number of classes.  

All techniques previously cited  determine a partition of the vertices
into classes. In  other words, every vertex is assumed  to belong to a
unique class.  This property may not correspond  to real applications,
in  which objects  often belong  to several  groups. Proteins  can for
instance  have more  than one  function  \citep{articlejeffery1999} or
scientists     belong     to     several    scientific     communities
\citep{articlepalla2005}. It is  therefore relevant to develop methods
in  order  to uncover  overlapping  structures  in  networks.  To  our
knowledge, the first
clustering  approach capable of  retrieving such  overlapping clusters
was  the  algorithm  of  \cite{articlepalla2005}  implemented  in  the
software CFinder \citep{manualpalla2006}.  For a given integer $k$, it
computes all the $k$-cliques  (complete subgraphs on $k$ vertices) and
all  the  pairs of  adjacent  $k$-cliques  ($k$-cliques sharing  $k-1$
vertices). A community is then defined as the vertex set of $k$-cliques which can be reached from each other through a sequence of adjacent $k$-cliques. Communities may then overlap without being merged if their intersection does not contain a $(k-1)$-clique. Decreasing the parameter $k$ leads to less cohesive but bigger communities. The choice of the optimal value for $k$ is then done heuristically by choosing the smallest value leading to no giant community. Moreover, this model can again only deal with assortative mixing. This is also the case for the more recent approaches of \cite{articleball2011} and \cite{proceedingsyang2013}, which both propose efficient methods to detect overlapping clusters in large networks, based on the assumption that relevant classes correspond to dense areas.

A first mixture based model with overlapping communities was proposed by \cite{articleairoldi2008} and successfully applied on real networks \citep{proceedingsairoldi2006,articleairoldi2007}. This model, called Mixed Membership Stochastic Blockmodel (MMSB), is an adaptation of earlier mixed membership models \citep{articleblei2003,proceedingsgriffiths2005} to the context of networks.      In MMSB, a mixing weight vector $\bpi_{i}$ is drawn
from  a  Dirichlet  distribution  for  each  vertex  in  the  network,
$\pi_{iq}$  being the  probability of  vertex $i$  to belong  to class
$q$. For each couple $(i,j)$, a vector $\bZ_{i \rightarrow j}$ is sampled from a multinomial distribution $\mathcal{M}(1, \: \bpi_{i})$ and describes the class membership of vertex $i$ in its relation towards vertex $j$.
The edge probability from vertex $i$ to vertex $j$ is then given by $p_{ij}=\bZ_{i \rightarrow j}^{\intercal}\bB\bZ_{i \leftarrow i}$, where $\bB$ is a $Q \times Q$ matrix of connection probabilities similar to the $\bPi$ matrix in SBM. The model parameters are estimated through variational techniques and the number of classes is selected by using a BIC criterion.
No assumption being made on the matrix $\bB$, this model is as flexible as SBM. 
Moreover, depending on its  relations with other vertices, each vertex
can belong  to different classes and  therefore MMSB can  be viewed as
allowing overlapping clusters. However, the limit of MMSB is that once
the vector $\bZ_{i \rightarrow j}$ has be drawn, the fact that $i$ may
belong to several classes, in its relations to other vertices, does not
influence the probability $p_{ij}$. Therefore, MMSB does not produce edges
which are themselves influenced by  the fact that some vertices belong
to multiple clusters.

 \cite{articlelatouche2011}  propose  another   extension  of  SBM  to
 overlapping classes, called the Overlapping Stochastic Block Model (OSBM). The main difference with SBM and MMSB is that the latent classes $\bZ$ are no longer drawn from multinomial distributions but from a product of Bernoulli distributions. In other words, to each vertex $i$ corresponds a $\{0-1\}$ vector $\bZ_i$ describing the classes it belongs to, and $\bZ_i$ may contain one,  several, or no coordinates equal to $1$. The connection probabilities are then determined by using a connectivity matrix like for SBM. The model parameters are estimated in a frequentist framework by using two successive approximations of the log-likelihood. Simulations show a better behaviour of this model for retrieving structures on a fixed number of classes in comparison with CFinder and MMSB. However, it suffers from a lack of criterion to choose the right number of classes.

Our main  concern in this paper  is to derive a  criterion to estimate
the number of classes in OSBM. To do so, we rely on the Bayesian framework and take advantage of the 
marginal likelihood  $p(\bX)$, which provides  a consistent estimation
of the distribution of the data \citep{articlebiernacki2010}.  Since the
marginal likelihood is not tractable directly in OSBM, we derive a non
asymptotic approximation which is obtained
using a variational Bayes EM algorithm.

In  Section \ref{sect:model},  we review  the OSBM  model  proposed by
\cite{articlelatouche2011}. Then, we introduce
conjugate  prior  distributions for  the  model  parameters.  In Section
\ref{sect:estim}, a variational Bayes EM algorithm is derived to perform inference
along   with   a  model   selection   criterion,  called   $IL_{osbm}$
(Integrated Likelihood for OSBM model), in Section
\ref{sect:modselec}.   Finally,   in  Section   \ref{sect:experiment},
experiments on simulated data and  on a subset of the French political
blogosphere network are carried  out.  Results illustrate the accuracy
of the  recovered clusters using the  overlapping clustering procedure
and show that
$IL_{osbm}$  is  a  relevant  criterion  to  estimate  the  number  of
overlapping clusters in networks.

\section{A Bayesian Overlapping Stochastic Block Model}\label{sect:model}

The data we model consists of  a $N \times N$ binary matrix $\bX$ with
entries $X_{ij}$  describing the presence  or absence of an  edge from
vertex $i$ to  vertex $j$. Both directed and  undirected relations can
be   analyzed   but  in   the   following,   we   focus  on   directed
relations. Moreover,  we assume  that the graph  we consider  does not
contain any self  loop. Therefore, the variables $X_{ii}$  will not be
taken into account. 

\subsection{Introducing the  Overlapping Stochastic Block Model}

The  Overlapping  Stochastic Block  Model  (OSBM)  associates to  each
vertex of a network a latent binary vector $\bZ_{i}=(Z_{iq})_{q=1\cdots Q}$ drawn from a multivariate Bernoulli distribution:
\begin{equation*}
  p(\bZ_{i} = \bz_{i} ) = \prod_{q=1}^{Q}\alpha_{q}^{z_{iq}}(1-\alpha_{q})^{1-z_{iq}},
\end{equation*}
where $Q$ denotes the number  of classes considered. Note that in this
model, each  vertex is not characterized  by one class  as in standard
mixture models.  Indeed, the  $\{0-1\}$ vector $\bZ_i$  indicating the
classes  of vertex  $i$ may  contain several  $1$'s, meaning  that the
vertex belongs to several classes.  It may also contain only $0$'s, so
that the corresponding vertex belongs to no class in the network. The latter
phenomenon may appear as a drawback  but is in fact an advantage of the
model as mixture models for networks, when applied to real data, often
show  one  heterogeneous  class  containing  all  vertices  with  weak
connection profiles  \citep{articledaudin2008}.  Rather than  using an
extra component to model these outliers, OSBM relies on the null component
such $\bZ_{i}=\mathbf{0}$ if vertex $i$ is an outlier and should not be
classified in any class. 
\smallskip

The edges are then assumed to be drawn from a Bernoulli distribution:
\begin{equation*}
  X_{ij}|\bZ_{i},\bZ_{j} \sim \Bernoulli\big(X_{ij};\:g(a_{\bZ_{i},\bZ_{j}})\big),
\end{equation*}
where 
\begin{equation*}
  a_{\bZ_{i}, \bZ_{j}} = \bZ_{i}^{\intercal} \bW \bZ_{j} +
  \bZ_{i}^{\intercal}\bU + \bV^{\intercal}\bZ_{j} + W^{*},
\end{equation*}
and   $g(x)  =   (1  +   e^{-x})^{-1}$  is   the   logistic  sigmoid
function.  The  first  term  in  the  right-hand  side  describes  the
interactions between vertices $i$ and $j$ using $\bW$ a $Q\times Q$
matrix. 
The second term parametrized by vector $\bU$ represents the
overall capacity of  vertex $i$ to emit edges  and, symmetrically, the
third term parametrized by vector $\bV$ represents the capacity of vertex $j$ to receive edges. Finally,
$ W^{*}$  is the parameter  controlling sparsity as $g(W^{*})$  is the
probability  to see  an  edge  between two  vertices  belonging to  no
class. 

Note that the use of the logistic function $g$ implies that
\begin{equation*}
  p(X_{ij}=x_{ij}|\bZ_{i},\bZ_{j}) = e^{x_{ij}a_{\bZ_{i}, \bZ_{j}}}g(-a_{\bZ_{i}, \bZ_{j}}),
\end{equation*}

Finally, to simplify notations, we define $\tilde{\bZ_{i}}     =    \big(\bZ_{i},     1\big)^{\intercal},\forall i$ and
\begin{displaymath}
  \tilde{\bW} = \begin{pmatrix}
    \bW & \bU \\
    \bV^{\intercal} & W^{*}
  \end{pmatrix},
\end{displaymath}
so that 
\begin{equation*}
  a_{\bZ_{i}, \bZ_{j}} = \tilde{\bZ_{i}}^{\intercal} \tilde{\bW} \tilde{\bZ_{j}}.
\end{equation*}

The latent variables $\bZ_{1},\dots,\bZ_{N}$ are iid and given this latent structure, all the edges are supposed to be independent.
 When considering a directed graph without self loops, conditional distributions can therefore be written as:
\begin{equation*}
 p(\bZ| \balpha) = \prod_{i=1}^{N} \prod_{q=1}^{Q}\alpha_{q}^{Z_{iq}} (1 - \alpha_{q})^{1 - Z_{iq}},
\end{equation*}
and
\begin{equation}\label{eq:conDist}
  \begin{aligned}
  p(\bX|\bZ, \tilde{\bW}) &= \prod_{i\neq j}^{N}p(X_{ij}|\bZ_{i},\bZ_{j},\btW) \\
  &=\prod_{i \neq j}^{N}  e^{X_{ij}a_{\bZ_{i}, \bZ_{j}}}g(-a_{\bZ_{i}, \bZ_{j}}). 
  \end{aligned}
\end{equation}

\subsection{Fitting OSBM into a Bayesian framework}

Let us now describe OSBM in a full Bayesian framework by introducing some conjugate prior distributions for the model parameters. Since $p(\bZ_{i}|\balpha)$ is a multivariate Bernoulli distribution, we consider independent Beta distributions for the class probabilities:
\begin{equation*}
  p(\balpha) = \prod_{q=1}^{Q} \Beta(\alpha_{q};\: \eta_{q}^{0},\zeta_{q}^{0}),
\end{equation*}
where $\eta_{q}^{0}=\zeta_{q}^{0}=1/2,\forall q$. This corresponds to a product of non-informative Jeffreys prior distributions. A uniform distribution can also be chosen simply by fixing $\eta_{q}^{0}=\zeta_{q}^{0}=1,\forall q$.

In order to model the $(Q+1) \times (Q+1)$  real matrix $\btW$, we consider the  $\asvec$  operator which stacks the columns of a matrix into a vector. Thus, if $\bA$ is a $2 \times 2$ matrix such that:
\begin{equation*}
  \bA = \begin{pmatrix}
    A_{11} & A_{12} \\
    A_{21} & A_{22}
  \end{pmatrix},
\end{equation*}
then 
\begin{equation*}
  \bA^{\asvec} = \begin{pmatrix}
    A_{11} \\ A_{21} \\
    A_{12} \\ A_{22}
    \end{pmatrix}.
\end{equation*}
Following the work of \cite{articlejaakkola2000} on Bayesian logistic regression, where an isotropic Gaussian distribution is used for the weight vector, we model the vector $\btW^{\asvec}$ using a multivariate Gaussian prior distribution with mean vector $\btW_{0}^{\asvec}$ and covariance matrix $\covmat_{0} = \frac{\bI}{\beta}$:
\begin{equation*}
  p(\btW^{\asvec}|\beta) = \Gaussian(\btW^{\asvec};\: \btW_{0}^{\asvec}, \frac{\bI}{\beta}).
\end{equation*}
We denote $\bI$ the identity matrix and in all the experiments that we
carried out, we set $\btW_{0}^{\asvec}=\mathbf{0}$. This approach can easily be extended to more general Gaussian priors by considering, for instance, a full covariance matrix $\covmat_{0}$ or by associating a different hyperparameter with different subsets of the parameters in $\btW$.

Finally, we consider a Gamma distribution to model the hyperparameter $\beta$:
\begin{equation*}
  p(\beta) = \Gam(\beta;\: a_{0},b_{0}).
\end{equation*}
By construction,  the Gamma distribution  is informative. In  order to
limit its influence on the  posterior distribution, a common choice in
the  literature is  to  set the  hyperparameters  $a_{0}$ and  $b_{0}$,
 controlling the  scale and rate  respectively, to low values.  In our
 experiments, we set $a_{0}=b_{0}=1$.

\section{Estimation}\label{sect:estim}

In this section, we propose a Variational Bayes EM (VBEM) algorithm, based on global and local variational techniques, which leads to an approximation of the full posterior distribution over the model parameters and latent variables, given the observed data $\bX$. This procedure relies on a lower bound which will be later used as non asymptotic approximation of the marginal log-likelihood $\log p(\bX)$.

\subsection{Variational approximation}

The integrated log-likelihood under the OSBM model can be written as:
\begin{equation*}
  \log p(\bX) = \sum_{\bZ} \int \int \int p(\bX|\bZ, \tilde{\bW}) p(\bZ| \balpha)  p(\btW^{\asvec}|\beta) p(\balpha) p(\beta) d\balpha d\btW d\beta.
\end{equation*}
However, as it is often the case when considering mixture models, the
exponential  number of terms  in the  summation makes  its computation
intractable.       The       well       known       EM       algorithm
\citep{articledempster1977,bookmclachlan1997}  cannot  be  applied  as
such to perform inference as it would require the posterior
distribution $p(\bZ|\bX, \balpha, \btW,\beta)$ to be tractable. Therefore, we propose to use a variational approximation, which relies on the decomposition of the marginal log-likelihood into two terms:
\begin{equation*}
  \log p(\bX) = \LowerBound(q) + \KL\left((q(\cdot) || p(\cdot|\bX)\right),
\end{equation*}
where
\begin{equation} \label{eq:lowerBound}
  \LowerBound(q) = \sum_{\bZ} \int \int \int q(\bZ, \balpha, \btW, \beta) \log \left\{\frac{p(\bX, \bZ, \balpha, \btW, \beta)}{q(\bZ, \balpha, \btW, \beta)}\right\} d\balpha d\btW d\beta, 
\end{equation}
and
\begin{equation} \label{eq:kullback}
  \KL(q(\cdot) || p(\cdot | \bX)) = -\sum_{\bZ} \int \int \int q(\bZ, \balpha, \btW, \beta) \log \left\{\frac{p(\bZ, \balpha, \btW, \beta |\bX)}{q(\bZ, \balpha, \btW, \beta)}\right\} d\balpha d\btW d\beta.
\end{equation}

$\LowerBound$   is    a   lower    bound   of   $\log    p(\bX)$   and
$\KL(\cdot||\cdot)$  denotes the  Kullback-Leibler  divergence between
the        distributions        $q(\bZ,\balpha,\btW,\beta)$        and
$p(\bZ,\balpha,\btW,\beta|\bX)$.   Note  that   when   $q(\cdot)$  and
$p(\cdot|\bX$) are  equal, the Kullback-Leibler  distance vanishes and
$\LowerBound(q)$  is  equal  to  the  integrated  log-likelihood.  The
maximization  of  $\LowerBound(q)$ and  the  minimization  of the  KL
divergence  are therefore equivalent problems.  

However, to obtain a tractable algorithm, two further approximations are  needed. First, the search space for the functional $q(\cdot)$ is limited to factorized distributions, that is we assume that $q(\bZ,\balpha,\btW,\beta)$ can be written as:
\begin{equation*}
  q(\bZ, \balpha, \btW,\beta) = q(\balpha)q(\btW)q(\beta)q(\bZ) = q(\balpha)q(\btW)q(\beta)\big(\prod_{i=1}^{N}\prod_{q=1}^{Q} q(Z_{iq})\big).
\end{equation*}

Second, the lower bound $\LowerBound$ is still intractable due to the logistic function in the distribution $p(\bX |\bZ,  \btW)$ (see Equation \ref{eq:conDist}). Therefore, we consider, for a given $N \times N$ positive real matrix $\bxi$, the tractable lower bound obtained by \cite{articlejaakkola2000}:

\begin{prop}\label{prop:bound}
  (Proof in Appendix \ref{sec:lowBound1})
  Given any $N \times N$ positive real matrix $\bxi$, a lower bound of the first lower bound is given by:
  \begin{equation*}
    \log p(\bX) \geq \LowerBound(q) \geq \LowerBound(q;\:\bxi),
  \end{equation*}
  where
  \begin{equation*}
    \LowerBound(q;\: \bxi) = \sum_{\bZ} \int \int \int q(\bZ, \balpha, \btW, \beta) \log\big(\frac{h(\bZ, \btW, \bxi)p(\bZ,\balpha,\btW,\beta)}{q(\bZ, \balpha, \btW, \beta)}\big) d\balpha d\btW d\beta,
  \end{equation*}
  and
  \begin{equation*}
     \log h(\bZ, \btW, \bxi) =  \sum_{i \neq j}^{N}\left\{(X_{ij}-\frac{1}{2})a_{\bZ_{i},\bZ_{j}} - \frac{\xi_{ij}}{2} + \log g(\xi_{ij}) - \lambda(\xi_{ij})(a_{\bZ_{i},\bZ_{j}}^{2} - \xi_{ij}^{2})\right\},
  \end{equation*}
where $\lambda(\xi) = (g(\xi)-1/2)/(2\xi)$.
\end{prop}

The lower bound $\log h(\bZ,  \btW, \bxi)$ of $\log p(\bX |\bZ, \btW)$
can  be  tight as  it  is obtained  through  a  Taylor expansion.  The
precision  of the approximation  obtained by  integrating it  over the
distributions of $\bZ$, $\balpha$, $\btW$ and $\beta$ cannot be evaluated but obviously depends on the choice of $\bxi$. We therefore propose an inference algorithm based on the alternate updating of the global variable set $\{\bZ, \btW, \balpha, \beta\}$ and the local parameter matrix $\bxi$.

\subsection{Variational Bayes EM}\label{ssec:vbem}

Suppose first that $\bxi$ is held fixed.
In    order     to    approximate    the     posterior    distribution
$p(\bZ,\balpha,\btW,\beta|\bX)$        with       a       distribution
$q(\bZ,\balpha,\btW,\beta)$, a VBEM algorithm \citep{proceedingsbeal03,articlelatouche2012} is applied on the lower bound $\LowerBound(q;\:\bxi)$. Such an algorithm mimics the classic EM algorithm by alternating an updating of the distribution $q(\bZ)$ (the variational E-step) and updating of the distributions $q(\btW)$, $q(\balpha)$ and $q(\beta)$ (variational M-step). The update of each of those distributions is done by integrating the lower bound with respect to all distributions but the one of interest. The functional forms of all the priors were chosen such that the updates generate distributions of the same functional form, so that only the value of the hyperparameters have to be changed.
This procedure ensures  the convergence of the algorithm to a local maximum of $\LowerBound(q;\:\bxi)$.

In   the  case   of   $q(\bZ)$,   the  updated   value   is  the   set
$(\tau_{iq})_{1\leq i\leq N,1\leq Q}$  which corresponds to the set of
(approximated) posterior probabilities for each individual to belong to each group.

The validity of this approach relies on the results of the following theorem:

\begin{thm}\label{thm:vbem}
Consider a variable $\mathbf{Y} \in \{\bZ,\btW,\balpha,\beta \}$ which distribution is of the same functional form the corresponding prior defined in Section~\ref{sect:model} and which depends on a set of hyperparamaters $\btheta^{0}$. Consider the updating of this variable by the VBEM algorithm. 

The obtained distribution is then  of the same functional form  as the prior and the new hyperparameter set $\btheta^{N}$ is obtained by applying the relevant formulae among the following:
\begin{description}
\item[for $\mathbf{Y}=\bZ$] 
\begin{equation*}
     \begin{aligned}
       \tau_{iq} &= g\bigg\{\psi(\eta_{q}^{N})-\psi(\zeta_{q}^{N}) + \sum_{j\neq i}^{N}(X_{ij}-\frac{1}{2})\bttau_{j}^{\intercal}(\btW_{N}^{\intercal})_{\cdot q} + \sum_{j \neq i}^{N}(X_{ji}-\frac{1}{2})\bttau_{j}^{\intercal}(\btW_{N})_{\cdot q}  \\
       &\:\:\:\:\: -\Trace\Big(\big(\bSigma_{qq}^{'} + 2\sum_{l\neq q}^{Q+1}\ttau_{il}\bSigma_{ql}^{'}\big)\big(\sum_{j\neq i}^{N}\lambda(\xi_{ij})\btE_{j}\big) + \big(\bSigma_{qq} + 2\sum_{l\neq q}^{Q+1} \ttau_{il}\bSigma_{ql}\big)\big(\sum_{j\neq i}^{N}\lambda(\xi_{ji})\btE_{j}\big)\Big)\bigg\},
     \end{aligned}
   \end{equation*}
   with $\bSigma_{ql}=\Esp_{\btW_{q},\btW_{l}}[\btW_{\cdot q} \btW_{\cdot l}^{\intercal}]$ and  $\bSigma_{ql}^{'} = \Esp_{\btW_{q\cdot},\btW_{l\cdot}}[\btW_{q\cdot}^{\intercal}\btW_{l\cdot}]$
\smallskip

\item[for $\mathbf{Y}=\btW$]
 \[
 \btW_{N}^{\asvec} =  \covmat_{N}\left\{\sum_{i \neq j}^{N} (X_{ij}-\frac{1}{2}) \bttau_{j}\otimes \bttau_{i}\right\}, 
 \]
 with $\covmat_{N}^{-1}   =  \frac{a_{N}}{b_{N}}\bI + 2\sum_{i \neq j}^{N} \lambda(\xi_{ij})(\btE_{j}\otimes \btE_{i})  $
\smallskip

\item[for $\mathbf{Y}=\balpha$]
  \[ \eta_{q}^{N}   =   \eta_{q}^{0} + \sum_{i=1}^{N} \tau_{iq} \hspace{1cm} \mbox{and} \hspace{1cm} \zeta_{q}^{N}   =   \zeta_{q}^{0} + N - \sum_{i=1}^{N}\tau_{iq} \]
\smallskip

\item[for $Y=\beta$]  
 \[ a_{N}  =  a_{0} + \frac{(Q+1)^{2}}{2} \hspace{1cm} \mbox{and} \hspace{1cm} b_{N}  =  b_{0} + \frac{1}{2}\Trace(S_{N}) + \frac{1}{2}(\btW_{N}^{\asvec})^{\intercal}\btW_{N}^{\asvec}\]
\end{description}

\end{thm}

The proofs of this statement for each of the distributions, as well as the definition of the quantities $\bttau$ and $\btE$ and of the function $\psi$ used to simplify the formulas, are detailed in the Appendix.

\subsection{Optimization of $\xi$}\label{ssec:optXi}

So far, we have seen how a VBEM algorithm could be used to obtain an approximation of the posterior distribution $p(\bZ,\balpha,\btW,\beta|\bX)$ for a given matrix $\bxi$. However, we have not addressed yet how $\bxi$ could be estimated from the data. We follow the work of \cite{proceedingsbishop2003} on Bayesian hierarchical mixture of experts. Thus, given a distribution $q(\bZ,\balpha,\btW,\beta)$, the lower bound $\LowerBound(q;\:\bxi)$ is maximized with respect to each variable $\xi_{ij}$ in order to obtain the tightest lower bound $\LowerBound(q;\:\bxi)$ of $\LowerBound(q)$. As shown in Proposition \ref{prop:optXi} and Appendix \ref{sec:optXi}, this optimization leads to  estimates $\hat{\xi}_{ij}$ of $\xi_{ij}$. 

\begin{prop}\label{prop:optXi}
  (Proof in Appendix \ref{sec:optXi}) An estimate $\hat{\xi_{ij}}$ of $\xi_{ij}$ is given by:
  \begin{equation*}
    \hat{\xi}_{ij} = \sqrt{\Trace\Big(\big(\covmat_{N} + \btW_{N}^{\asvec}(\btW_{N}^{\asvec})^{\intercal}\big)(\btE_{j}\otimes \btE_{i})\Big)}.
  \end{equation*}
\end{prop}

This gives rise to a three step optimization algorithm. Given a matrix $\bxi$, the variational Bayes E and M steps are used to approximate the posterior distribution over the model parameters and latent variables. The distribution $q(\bZ,\balpha,\btW,\beta)$  is then held fixed while the lower bound $\LowerBound(q;\:\bxi)$ is maximized with respect to $\bxi$. 
  These three stages are repeated until convergence of the lower bound
  (see  Algorithm \ref{algo:vbemOSBM}).  The distribution  $q(\bZ)$ is
  initialized       using      a       kmeans       algorithm.  

For all the experiments that we carried out, we set $\xi_{ij}=0.001,\forall i \neq j$. The computational cost of the algorithm is equal to $O(N^{2}Q^{4})$. The code, written in R, is available upon request.

\begin{algorithm}

  Initialize $\btau$ with a kmeans algorithm\;
  Initialize $\xi_{ij},\forall i \neq j$; $a_{N}=a_{0}$, $b_{N}=b_{0}$\;
 
  \Repeat{$\LowerBound(q;\:\bxi)$ converges}{
    
    $\btE_{i}\leftarrow \Esp_{\bZ_{i}}[\btZ_{i}\btZ_{i}^{\intercal}],\forall i$\;

    $\eta_{q}^{N} \leftarrow \eta_{q}^{0} + \sum_{i=1}^{N}\tau_{iq},\forall q$\;
    $\zeta_{q}^{N}     \leftarrow      \zeta_{q}^{0}     +     N     -
    \sum_{i=1}^{N}\tau_{iq},\forall q$\;
     $\covmat_{N}^{-1} \leftarrow \frac{a_{N}}{b_{N}}\mathbf{I} + 2\sum_{i \neq j}^{N} \lambda(\xi_{ij})(\btE_{j}\otimes \btE_{i})$\;

     $\btW_{N}^{\asvec}   \leftarrow   \covmat_{N}\left\{\sum_{i  \neq
         j}^{N}         (X_{ij}-\frac{1}{2})         \bttau_{j}\otimes
       \bttau_{i}\right\}$\;
 $a_{N}\leftarrow a_{0} + (1/2)(Q+1)^{2}$\;
    $b_{N}\leftarrow b_{0}+(1/2)\big(\Trace(\covmat_{N})+(\btW_{N}^{\asvec})^{\intercal}\btW_{N}\big)$\;
 
    $\xi_{ij} \leftarrow \sqrt{\Trace\Big(\big(\covmat_{N} + \btW_{N}^{\asvec}(\btW_{N}^{\asvec})^{\intercal}\big)(\btE_{j}\otimes \btE_{i})\Big)},\forall i \neq j$\;
   
    \Repeat{$\btau$ converges}{
      Compute $\tau_{iq},\forall (i,q)$ using Theorem \ref{thm:vbem}\;
     }
  }  
  \caption{Variational Bayes inference for overlapping stochastic block model when applied on a directed graph without self loop.}
\label{algo:vbemOSBM}
\end{algorithm}

\newpage
  
\section{Model Selection}\label{sect:modselec}

So  far,  the  number  of  latent  clusters has  been  assumed  to  be
known.  Given  $Q$,  we  showed  in  Section  \ref{ssec:vbem}  how  an
approximation of the posterior  distribution over the latent structure
and model parameters  could be obtained. We now  address the problem of
estimating the number of clusters directly from the data. 
Given a set of values of $Q$, we aim at selecting $Q^{*}$ which maximizes the marginal log-likelihood $\log p(\bX| Q)$, also called integrated observed-data log-likelihood. Unfortunately, this quantity is not tractable since for each value of $Q$, it involves integrating over all possible model parameters and latent variables:
\begin{equation*}
  \log   p(\bX|Q)  =   \log\left\{\sum_{\bZ}\int   \int  \int   p(\bX,
    \bZ,\balpha,\btW,\beta|Q)d\balpha d\btW d\beta\right\}.
\end{equation*}
We propose to replace the marginal log-likelihood with its variational
approximation. Thus, for  each value  of $Q$  considered, Algorithm
\ref{algo:vbemOSBM}    is    applied     in    order    to    maximize
$\LowerBound(q;\:\bxi)$ with  respect to $q(\cdot)$  and $\bxi$. After
convergence, the  lower bound is then  used as an  estimation of $\log
p(\bX|Q)$ and $Q^{*}$ is chosen such that the lower bound is maximized.  Obviously, this approximation cannot be verified
analytically because neither $\LowerBound(q)$ in (\ref{eq:lowerBound})
nor  the   Kullback-Leibler  divergence  in   (\ref{eq:kullback})  are
tractable. Nevertheless, we rely on such approximation, as in
  \cite{bookbishop2006,inbooklatouche2009,articlelatouche2012}, to propose a
tractable model selection criterion that we call $IL_{osbm}$.  We
  prove in the appendix (Appendix \ref{sec:lowProof}) that if computed right after the M step of the
  variational Bayes EM algorithm, the lower bound has the following expression:
\begin{multline*}\label{eq:ilosbm}
IL_{osbm} = \sum_{i\neq j}^{N} \left\{\log g(\xi_{ij}) - \frac{\xi_{ij}}{2} + \lambda(\xi_{ij})\xi_{ij}^{2}\right\} + \sum_{q=1}^{Q}\log \bigg\{\frac{\Gamma(\eta_{q}^{0}+\zeta_{q}^{0})\Gamma(\eta_{q}^{N})\Gamma(\zeta_{q}^{N})}{\Gamma(\eta_{q}^{0})\Gamma(\zeta_{q}^{0})\Gamma(\eta_{q}^{N}+\zeta_{q}^{N})}\bigg\} \\+ \log \frac{\Gamma(a_{N})}{\Gamma(a_{0})} + a_{0}\log b_{0} 
+ a_{N}(1 - \frac{b_{0}}{b_{N}} - \log b_{N}) + \frac{1}{2}(\btW_{N}^{\asvec})^{\intercal}\covmat_{N}^{-1}\btW_{N}^{\intercal} + \frac{1}{2}\log |\covmat_{N}|\\ - \sum_{i=1}^{N}\sum_{q=1}^{Q}\left\{\tau_{iq}\log \tau_{iq} + (1-\tau_{iq})\log(1-\tau_{iq})\right\},
\end{multline*}
where $\Gamma(\cdot)$ is the gamma function. We emphasize that $IL_{osbm}$ is the first model selection criterion
to be derived for OSBM.

\section{Experiment}\label{sect:experiment}

We recall that in \cite{articlelatouche2011}, we first
introduced the OSBM model along with a variational EM
algorithm. We gave an extensive series of comparison of
this  approach to other  widely used  graph clustering  methods. In
particular, OSBM was compared to  the (non overlapping) SBM model, the
MMSB     model    of    \cite{articleairoldi2008},     and    CFinder
\citep{articlepalla2005}.  This  set  of  experiments  illustrated  the
 capacity of  OSBM along with  the variational inference  algorithm to
 uncover overlapping clusters in networks. In light of these results, we now focus in this paper on the OSBM
model and we aim at evaluating our new contribution, \emph{i.e.} a
model selection criterion for OSBM. 

However,  because the quality  of the  inference procedure  we propose
obviously  depends  on the  variational  bounds,  we start  by
evaluating  the  approximations,   at  the  parameter  level,  through
credibility intervals  and a series of experiments on simulated data. Then, we illustrate the capacity of $IL_{osbm}$
 to retrieve the true number of clusters and evaluate
the accuracy of the recovered clusters. Finally, we apply our
methodology  to study  a subset  of the  French  political blogosphere
network (see \citep{articlezanghi2008}) and we
analyze the results, from the estimation of the number of clusters to
the clustering of the vertices.


\subsection{Simulated data}

The OSBM model is used in this set of experiments to generate networks
with  community structure, where  vertices of  a community  are mostly
connected to vertices of the  same community. 
%

To limit the number of free parameters, we consider the $Q \times Q$ real matrix $\bW$:
\begin{displaymath}
  \bW = \begin{pmatrix}
    \boldsymbol{\lambda} & -\epsilon & \dots & -\epsilon \\
    -\epsilon & \boldsymbol{\lambda} & & \vdots \\
     \vdots & & \ddots & -\epsilon  \\
     -\epsilon & \dots & -\epsilon & \boldsymbol{\lambda} \\
  \end{pmatrix},
\end{displaymath}
and the $Q$-dimensional real vectors $\bU$ and $\bV$:
\begin{displaymath}
  \bU = \bV = \begin{pmatrix}
     \epsilon & \dots & \epsilon \\
  \end{pmatrix}.
\end{displaymath}

\subsubsection{Variational Bayes credibility intervals}

\begin{figure}[!] \centering
  \setlength{\unitlength}{5mm}
  \includegraphics[width=8.5cm]{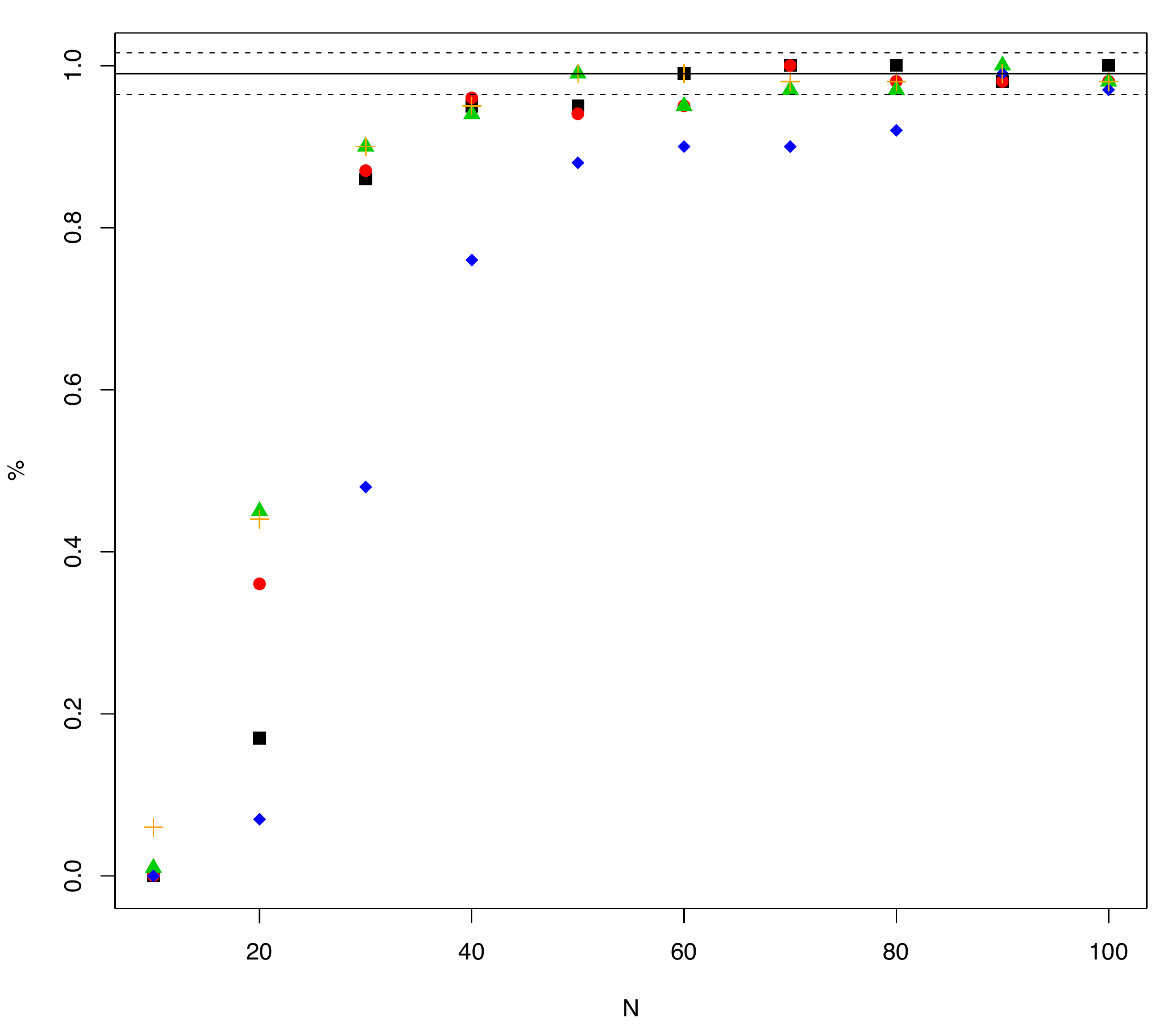}
  \caption{Proportions  of  the  simulations where  $99\%$  credibility
    intervals obtained with the  VBEM algorithm contain the true value
    of the parameters, for various values of $N$ (number of vertices).  $W_{11}$,
    black  square;  $W_{12}$,  red  circle; $U_{1}$,  green  triangle;
    $W^{*}$,  blue   diamond;  $\alpha_{1}$,  orange   cross.  Nominal
    credibility  ($99\%$), solid  line; binomial  confidence interval,
    dotted lines.}
  \label{fig:credInter}
\end{figure}

The   approximation   of  the   posterior   distribution  allows   the
construction  of   (approximate)  credibility  intervals.   Therefore,
following the
work of \cite{articlegazal2011} on the standard SBM model, we evaluate
here the inference of the model parameters that can
be  obtained  with the  VBEM  algorithm,  through  the quality  of  the
 credibility intervals estimated.  Thus, setting $\lambda=1.5$,
 $\epsilon=1$, and $W^{*}=-2$,  we  simulate  100  networks  with  $Q=3$
 classes,         having          the         same         proportions
 $\alpha_{1}=\alpha_{2}=\alpha_{3}=1/Q$, for various numbers $N$ of vertices
 in $\{10, 20,  \dots, 100\}$. For each network  generated, we run the
 VBEM algorithm with $Q=3$ classes and we calculate the proportions
 of credibility  intervals obtained containing  the true value  of the
 parameters.  Such  proportions should  present  binomial
 fluctuations    around   the    nominal   credibility.    In   Figure
 \ref{fig:credInter}, we  present the results  for $W_{11}$, $W_{12}$,
 $U_{1}$, $W^{*}$, and $\alpha_{1}$, for $99\%$ credibility intervals.
We observe that the actual credibility of the estimated intervals is
close to the nominal one as long as the network contains 80 vertices.

\subsubsection{Model selection and cluster assessment}

\begin{figure}[!] \centering
  \setlength{\unitlength}{5mm}
  \includegraphics[width=5cm]{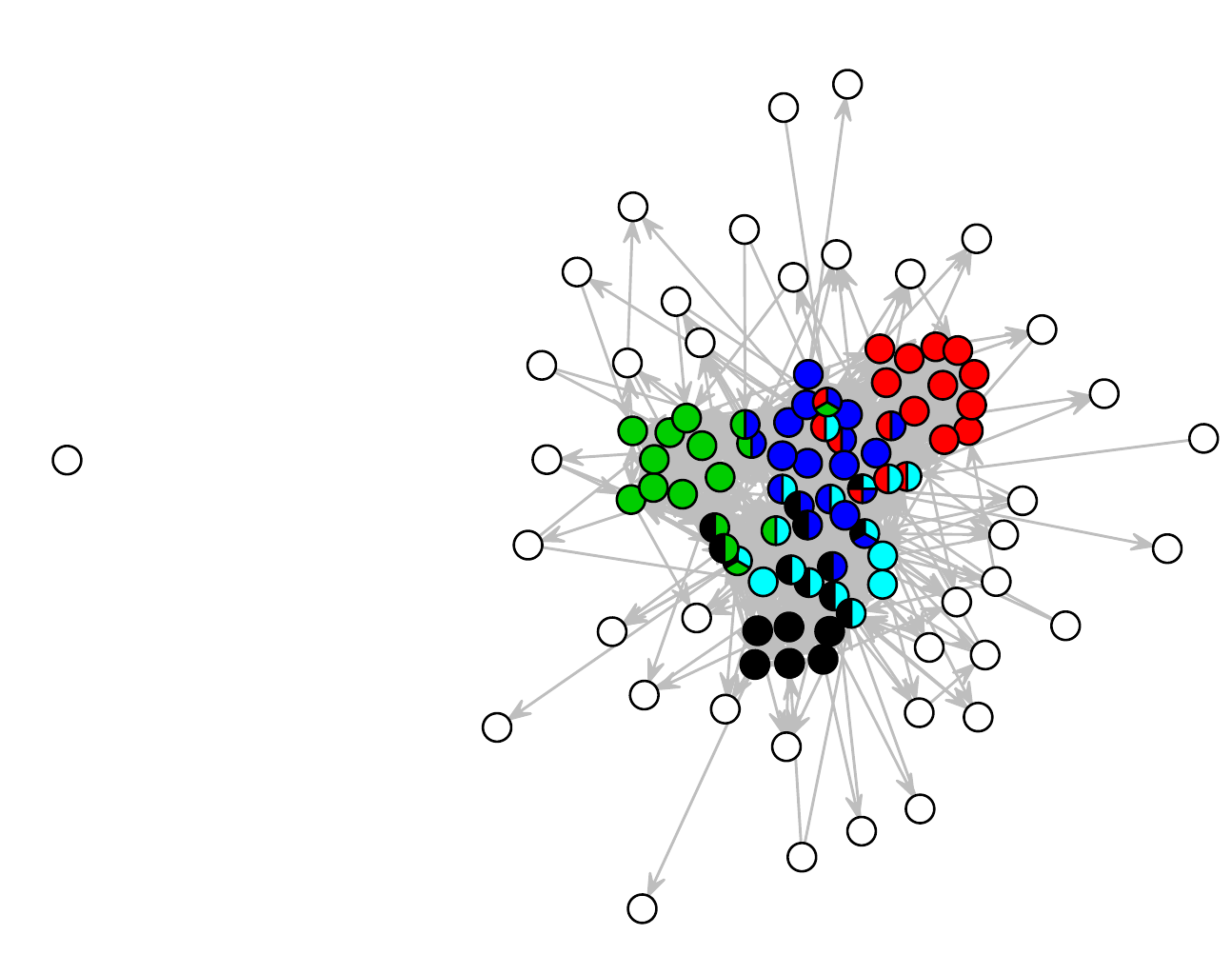}
  \caption{   Example    of   a   OSBM    network   with   $\lambda=6$
    ($p_{intra}\approx 0.9$) , $\epsilon=1$, $W^{*}=-5.5$, and $Q=5$ classes. Overlaps are represented using pies and outliers are in white.}
  \label{fig:aff1}
\end{figure}

We now aim at evaluating  $IL_{osbm}$ and the quality of the recovered
clusters. We first set $\epsilon=1$ and $W^{*}=-5.5$ which induces a probability
$p_{inter}=g(-\epsilon + 2 \epsilon + W^{*})\approx0.01$ of connection between each pair of vertices from
different    clusters.
   The values $\lambda\in\{6,4,3.5\}$ are then 
experimented. The corresponding probabilities $p_{intra}=g(\lambda + 2\epsilon +
W^{*})$ of connection between  vertices of the same
cluster  are  approximately  $0.9$,  $0.6$ and  $0.5$.   Moreover,  we
consider   various  numbers   $Q_{True}$  of   clusters  in   the  set
$\{2,\dots,7\}$ to generate networks, along with two scenarios depending on the type of
vector $\balpha$ considered. The {\em balanced  groups} correspond to
equal proportions $\alpha_{1}=\dots=\alpha_{Q_{True}}=1/Q_{True}$. The {\em unbalanced groups} correspond to groups of geometric size, that is
$\alpha_q \propto a^q$ and $\sum_{q=1}^{Q_{True}}\alpha_{q}=1$, for a=0.7.  Considering for example $Q_{True}=7$ produces
a  highly unbalanced  $\balpha=\{0.33, 0.23,  0.16, 0.11,  0.08, 0.05,
0.04\}$. Please note that the value of $a=0.7$ corresponds to an extreme case
scenario  which ensures,  with  a 0.99  chance  probability, that  the
smallest   class  has   at   least  one   element.   Thus,  for   each
$Q_{True}$, each $\lambda$, and
each type of vector $\balpha$,  we generate $100$  networks (see an example in Figure
\ref{fig:aff1}) with  $N=100$ vertices. 
\smallskip

The VBEM algorithm is
then applied  on each  network for various  numbers of classes  $Q \in
\{2,\dots,8\}$.          Note          that         we          choose
$\eta_{q}^{0}=1/2=\zeta_{q}^{0},\forall    q$,   $a_{0}=b_{0}=1$   and
$\btW^{vec}_{0}=\mathbf{0}$ for the hyperparameters. Like any optimization method, the overlapping clustering algorithm we propose depends on the initialization. Thus, for each simulated network and each number of classes $Q$, we consider $100$ initializations of $\btau$. Finally, we select the best learnt model for which the criterion $IL_{osbm}$ is maximized.
\smallskip

Two types of  outputs are generated to present  the results. The first
aims at  describing the  accuracy of the  recovered clusters. In order to
compare a true $\bZ$  with an estimated clustering matrix $\hat{\bZ}$,
we consider an index similar to the one proposed by \cite{proceedingsheller2007,proceedingsheller2008}:
\[ \sqrt{\frac{1}{N(N-1)} \sum_{i\neq j} | ( \bZ \bZ^{\intercal})_{ij}
  - (\hat{\bZ} \hat{\bZ}^{\intercal})_{ij} |} . \]
This can  be seen  as a  root mean square  error between 
$\bZ\bZ^{\intercal}$  and $\hat{\bZ}\hat{\bZ}^{\intercal}$.  These two
$N \times N$ matrices are invariant to column permutations of $\bZ$ and $\hat{\bZ}$
and  compute  the number  of  shared  clusters  between each  pair  of
vertices of a network. The better the classification, the lower this index, a null index indicating a perfect classification. The results associated to the $100$ generated networks are then summarized as boxplots.

The second type of results we generate is a confusion matrix which aims at showing the accuracy of the $IL_{osbm}$ criterion. It indicates both the real number of classes $Q_{True}$ and the number of classes selected by $IL_{osbm}$, the counts on the first diagonal corresponding to correct decisions.  
\smallskip

\begin{figure}[!] 
\centering
\caption{\label{fig:accuracyboxplots} 
Boxplots   of   $\sqrt{\frac{1}{N(N-1)}   \sum_{i\neq   j}  |(      \bZ
  \bZ^{\intercal})_{ij} - (\hat{\bZ} \hat{\bZ}^{\intercal})_{ij} |} 
\ $ for true number of classes $Q_{True}\in\{2,\dots,7\}$
   computed for balanced and unbalanced groups with three different $\lambda$ settings, $\lambda \in \{6,4,3.5\}$
      corresponding respectively to three different probabilities of intra group
      connection  $p_{intra}  \approx  \{0.9,0.6,0.5\}$.  Notice  that
      other parameters of simulation where set to $\epsilon=1$, $W^{*}=-5.5$ }
\begin{tabular}[htbp]{c c c}
& balanced groups & unbalanced groups\\
 \rotatebox{90}{\hspace{0.2cm } High connectivity ($\lambda=6$)} & \includegraphics[width=5cm]{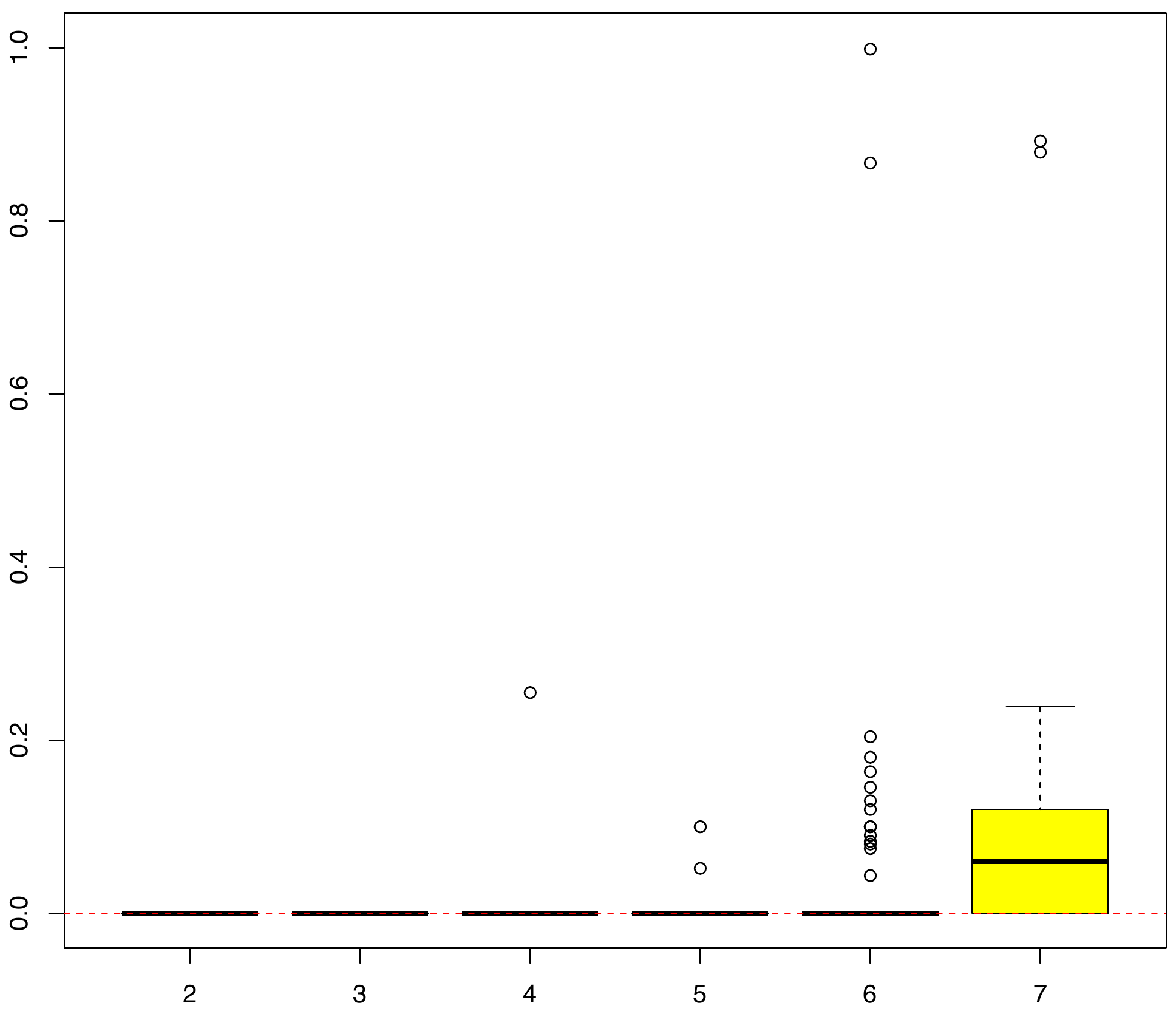}
       & \includegraphics[width=5cm]{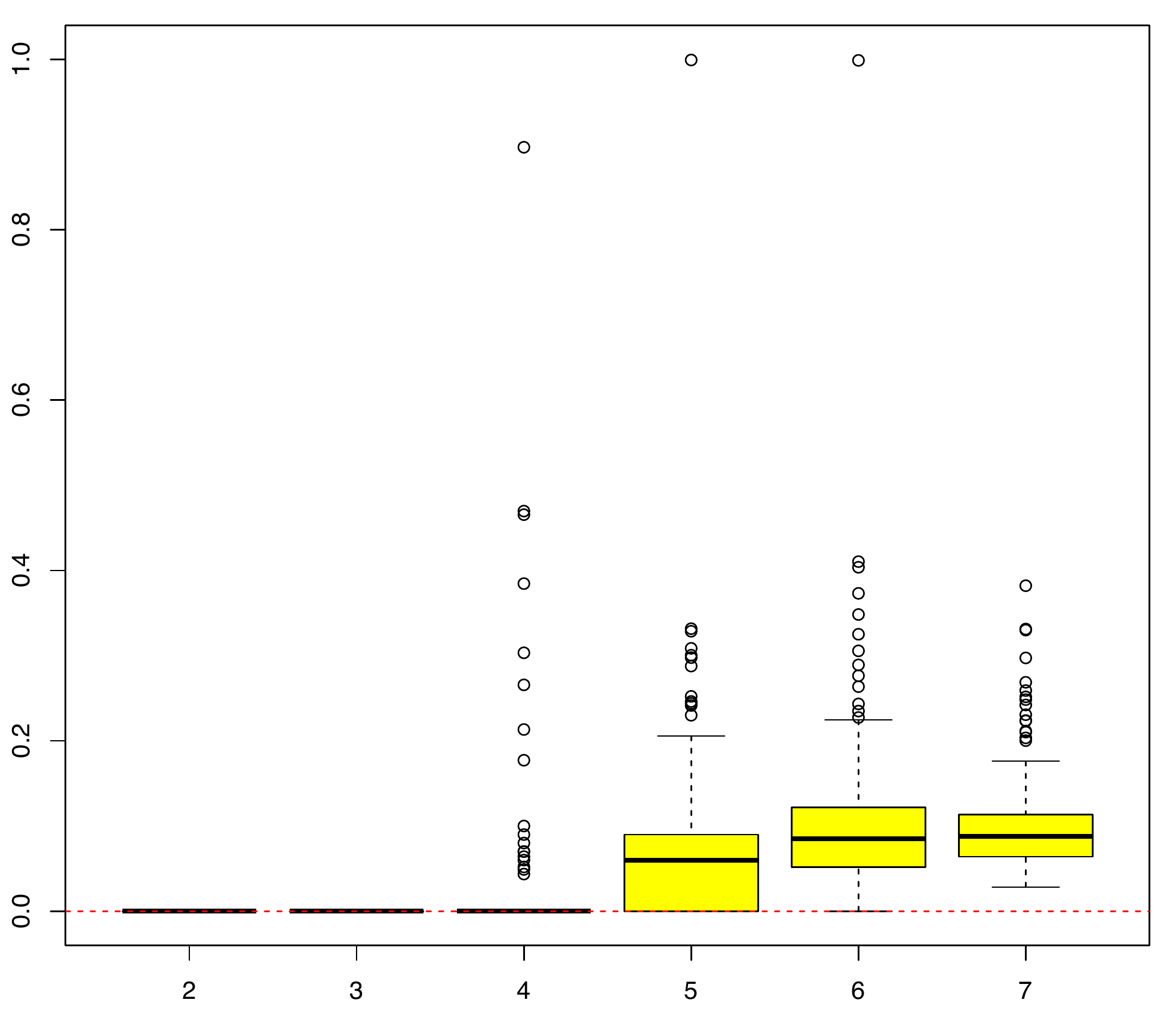} \\
\rotatebox{90}{\hspace{0.2cm } Average connectivity ($\lambda=4$)} & \includegraphics[width=5cm]{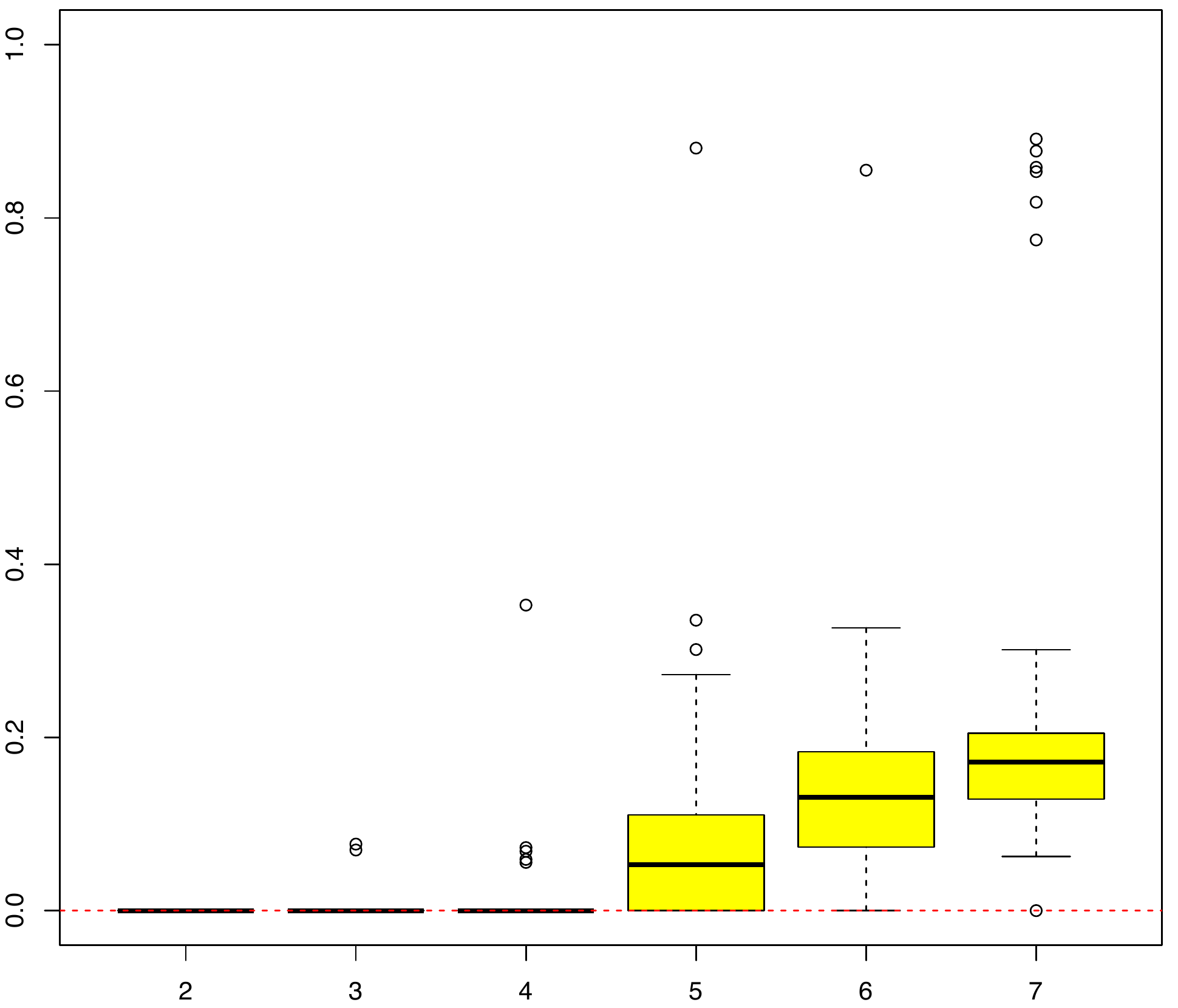}
       & \includegraphics[width=5cm]{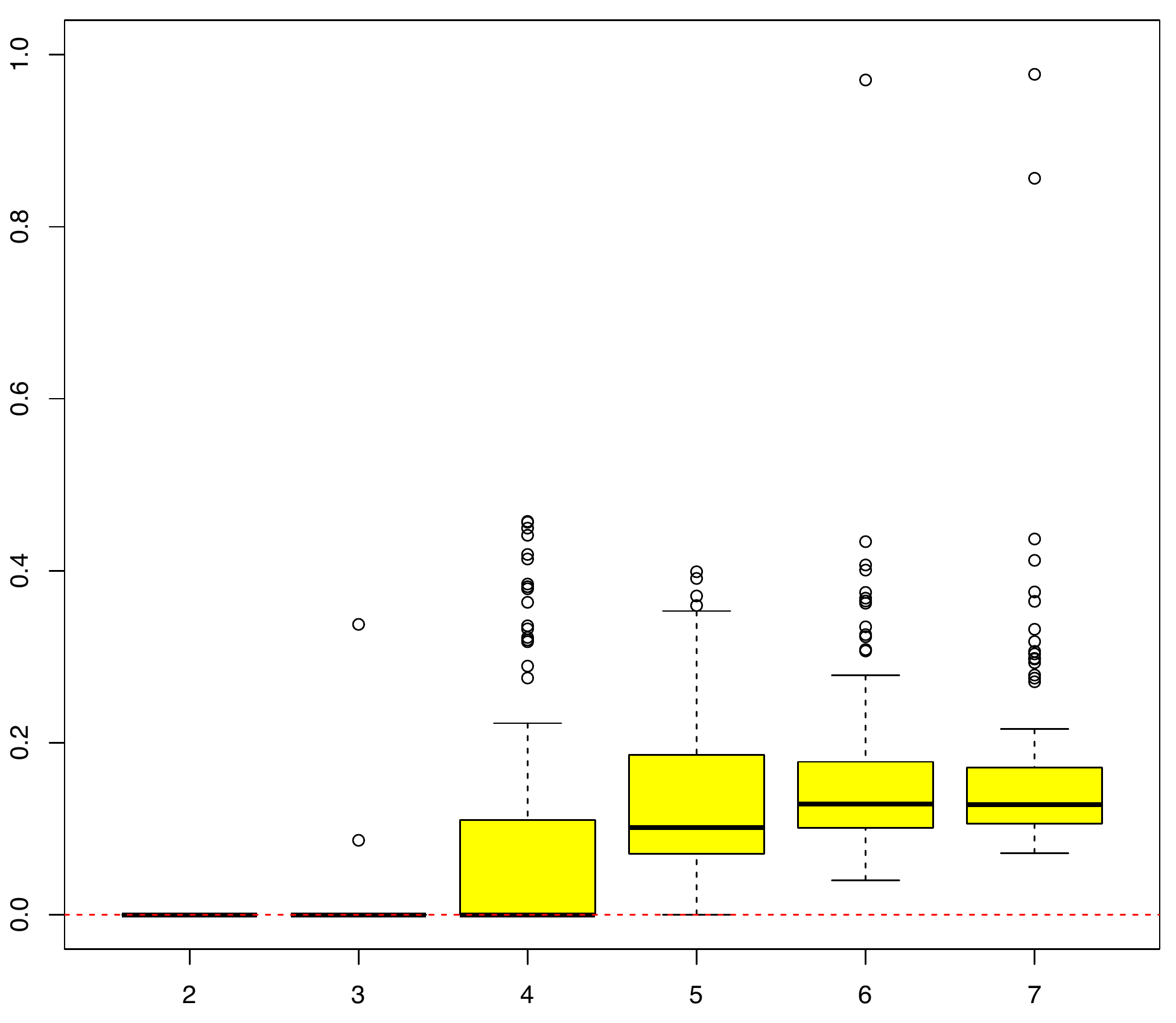}  \\
\rotatebox{90}{\hspace{0.2cm } Low connectivity ($\lambda=3.5$)} &    \includegraphics[width=5cm]{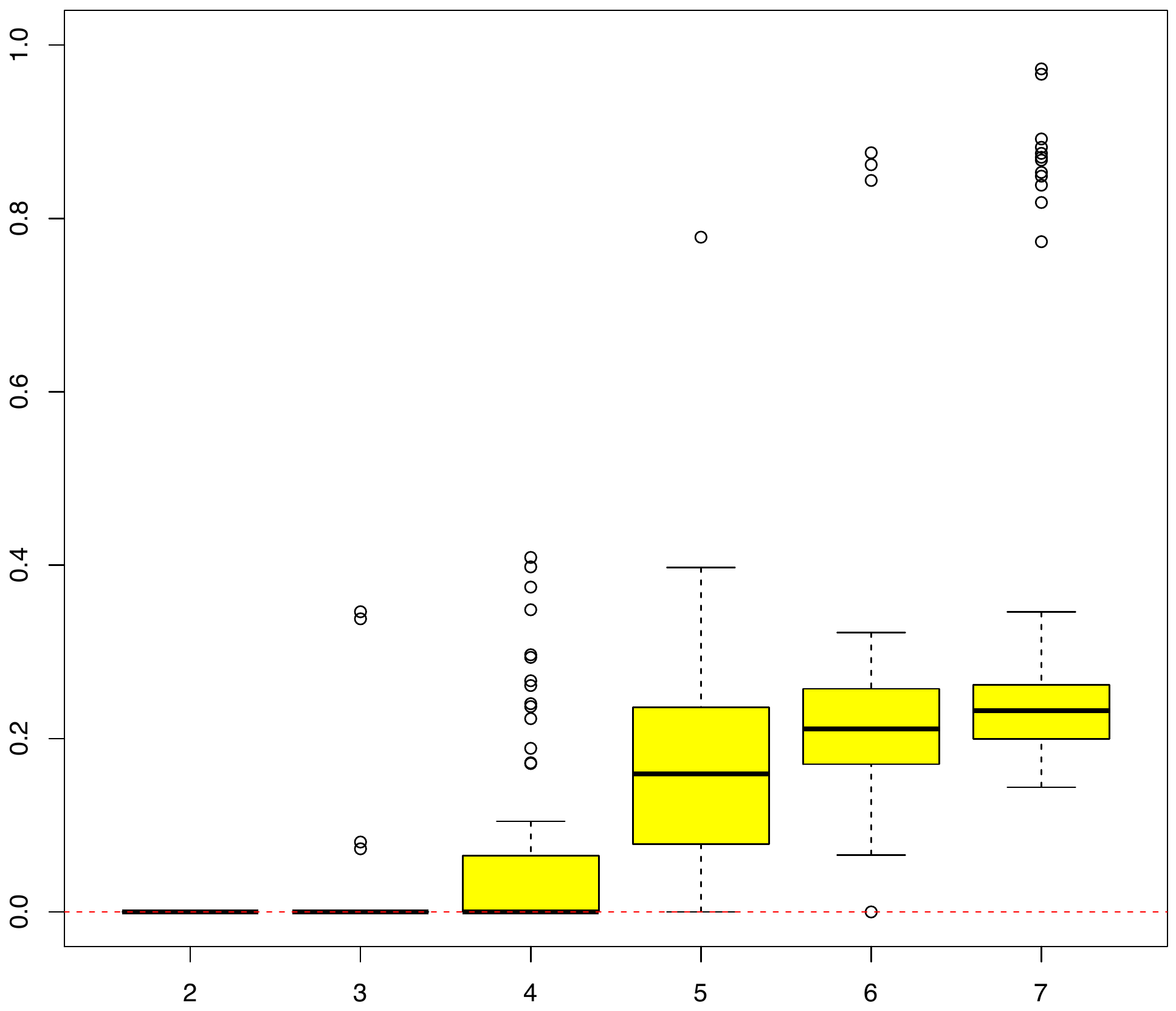}       & \includegraphics[width=5cm]{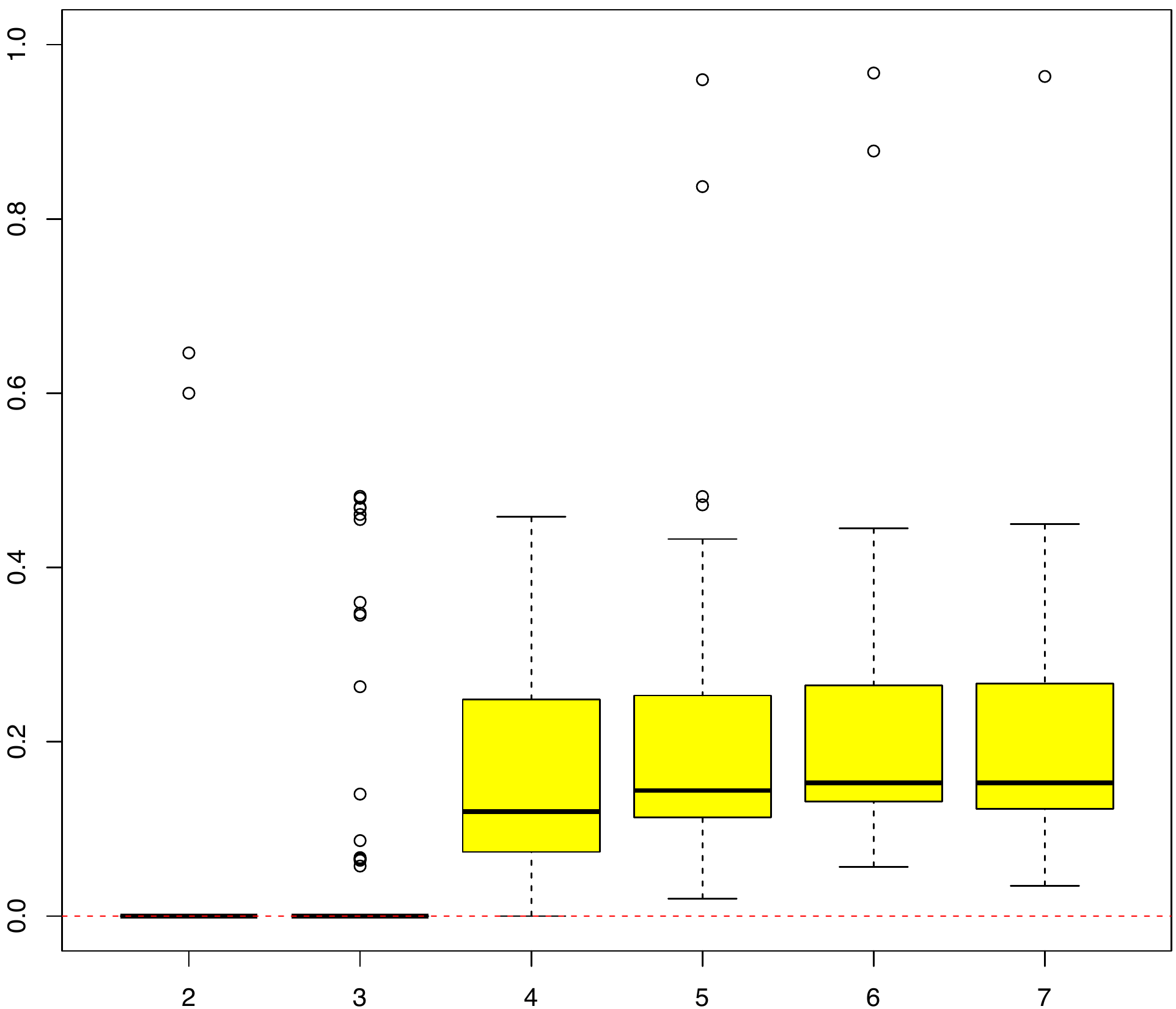}  
\end{tabular}
\end{figure}

The results are presented in Table \ref{table:confusionmatrices} and Figure \ref{fig:accuracyboxplots}. They
 illustrate 
the relevance of $IL_{osbm}$ criterion for estimating the number of
overlapping classes in networks, and  show that the OSBM learnt groups
are accurate. 
This is clearly the case when the graph is dense and the number of groups is low.
The performance of the model choice criterion decreases when the density
 within groups decreases, and when the balance between group proportions changes
(see Table \ref{table:confusionmatrices}). The
same behaviour  is observed concerning the quality  of the overlapping
clustering (see Figure \ref{fig:accuracyboxplots}). 

Let us consider for instance the balanced case with highly connected groups.
In that particular setting when 
$Q_{True}\in\{2,3\}$, $IL_{osbm}$ correctly estimate the number of
overlapping classes of $100$ out of the $100$ networks generated. For
$Q_{True}=5$, $IL_{osbm}$ still has $98$ percent accuracy.  The results then
  slowly deteriorate for $Q_{True} \in\{6,7\}$. Indeed, as $Q_{True}$
increases while the number of vertices remains unchanged, less  vertices  are  associated to  each  cluster  and
therefore it becomes more difficult to retrieve and distinguish the
overlapping communities.

The results obtained for the unbalanced setting follow the same pattern. They
are  just degraded compared  to balanced  setting and  tend to  show an
under-estimation of the number of groups when the connectivity within
groups   decreases    and   when    the   number   of    true   groups
increases.   Considering    for   example   $Q_{True}=2$    and   $\lambda=3.5$
($p_{intra}=0.5$),  the estimated  number of  classes is  accurate 98
times out of 100, but when $Q_{True}=7$ the estimated number of classes is 
correctly estimated  only 3 times out  of 100. Most of the time the model
choice strategy we proposed estimates 4
groups instead of 7. This demonstrates that correctly estimating the true number
of overlapping classes depends  both on the intra-connectivity and the
group balance. 

Considering  the   ability  of  OSBM  to   estimate  the  overlapping
communities,   the  boxplots   of   Figure  \ref{fig:accuracyboxplots}
exhibits   a   near   perfect    behaviour   in   all   settings   for
$Q_{True}\in\{2,3\}$. 
The  quality  of  these   performances  degrade  with  a  decrease  of
connectivity  as  well  as   a  difference  of  balance  between  group
proportion.



\begin{table}[htbp]
\caption{
\label{table:confusionmatrices} Confusion matrices for estimated number of
classes (columns ) $Q_{IL_{osbm}}\in\{2,\dots,8\}$ versus true
number of classes (rows) $Q_{True} \in\{2,\dots,7\}$ 
   computed for balanced and unbalanced groups with three different $\lambda$ settings, $\lambda \in \{6,4,3.5\}$
      corresponding respectively to three different probabilities of intra group
      connection  $p_{intra}  \approx  \{0.9,0.6,0.5\}$.  Notice  that
      other parameters of simulation where set to $\epsilon=1$, $W^{*}=-5.5$ }
\centering
\begin{tabular}{ c c  c c c c c c c | c c c c c c c}
&\multicolumn{1}{ c}{} & \multicolumn{7}{ c}{balanced groups} & \multicolumn{7}{c}{unbalanced groups} \\ 
& \multicolumn{1}{c}{} & \multicolumn{1}{c}{2} & 3 & 4 & 5 & 6 & 7 & \multicolumn{1}{c}{8} & 2 & 3 & 4 & 5 & 6
& 7 & \multicolumn{1}{c}{8} \\ \cline{3-16}
 \multirow{4}{*}{\rotatebox{90}{$\lambda=6$} } 
& 2 & $\mathbf{100}$ & 0 & 0 & 0 & 0 & 0 &0& $\mathbf{100}$ & 0 & 0 & 0 & 0 & 0 & 0 \\
& 3 & 0 & $\mathbf{100}$ & 0 & 0 & 0 & 0 & &0 &$\mathbf{100}$ & 0 & 0 & 0 & 0 & 0 \\
& 4 & 0 & 0 & $\mathbf{99}$ & 0 & 1 & 0 & 0 &0 & 6 & $\mathbf{85}$ & 5 & 3 & 1 & 0 \\
& 5 & 0 & 0 & 2 & $\mathbf{98}$ & 0 & 0 & 0 &0 & 3 & 34 & $\mathbf{50}$ & 8 & 4 & 1 \\
& 6 & 0 & 0 & 0 & 8 & $\mathbf{85}$ & 6 & 1 &0 & 0 & 29 & 49 & $\mathbf{15}$ & 6 & 1 \\
& 7 & 0 & 0 & 0 & 1 & 24 & $\mathbf{56}$ & 19 & 0 & 0 & 30 & 50 & 13 &
$\mathbf{6}$ & 1 \\  \cline{3-16}
 \multirow{4}{*}{\rotatebox{90}{$\lambda=4$} } 
&2 & $\mathbf{100}$ & 0 & 0 & 0 & 0 & 0 & 0& $\mathbf{100}$ & 0&  0&  0&  0 &0 &0\\
&3 & 0 & $\mathbf{100}$ & 0 & 0 & 0 & 0 & 0&   0& $\mathbf{99}$ & 1 & 0 & 0 &0 &0\\
&4 & 0 & 0 & $\mathbf{99}$ & 1 & 0 & 0 & 0 &   0 &14 &$\mathbf{68}$ & 9 & 7 &2 &0\\
&5 & 0 & 0 & 4 & $\mathbf{79}$ & 14 & 1 & 2&   0 &18 &50& $\mathbf{22}$&  4& 6 &0\\
&6 & 0 & 0 & 1 & 22 & $\mathbf{49}$ & 22 &6&   0 &20& 46& 16& $\mathbf{13}$& 4& 1\\
&7 & 0 & 0 & 0 & 16 & 47 & $\mathbf{24}$&13& 0 &22& 56& 14& 5& $\mathbf{3}$& 0\\  \cline{3-16}
 \multirow{4}{*}{\rotatebox{90}{$\lambda=3.5$} } 
& 2 &$\mathbf{100}$ &0 &0& 0 &0 &0 &0 &$\mathbf{98}$ &2& 0& 0& 0& 0& 0\\
&3  & 0 &$\mathbf{98}$  & 2 & 0 & 0 & 0 & 0 & 1 &$\mathbf{91}$ & 7 & 0 &1 &0& 0\\
&4 &  0 & 0& $\mathbf{87}$&  9&  3&  1&  0&  1& 43 &$\mathbf{32}$ &16& 4 &1 &3\\
&5 &  0 & 0 &15 &$\mathbf{44}$ &26& 12&  3&  2& 34&44&  $\mathbf{9}$& 8& 3& 0\\
&6 &  0 & 1 &11 &28 &$\mathbf{22}$& 25& 13&  0 &47& 32& 15& $\mathbf{5}$ &1 &0\\
&7 &  0&  0&  6& 34& 28& $\mathbf{17}$ &15 & 2 &30 &46 &14 &5 &$\mathbf{3}$& 0\\
\end{tabular}
\end{table}

\subsection{French political blogosphere}

As in \cite{articlelatouche2011}, we consider a subset of the French
political  blogosphere network. The  network is  made of  196 vertices
connected by  2864 edges. It was  built from a single  day snapshot of
political  blogs  automatically extracted  on  14th  october 2006  and
manually classified  by the ``Observatoire  Pr\'{e}sidentiel Project''
\citep{articlezanghi2008}.   Vertices correspond to hostnames and there is
an  edge between  two vertices  if there  is a  known hyperlink  from one
hostname to another. The five main political parties which are present
in the  data set  are the UMP  (french ``republican''),  liberal party
(supporters of economic-liberalism), UDF (``moderate''
party), PRG  (``extreme left wing'') and PS  (french ``democrat''). In
\cite{articlelatouche2011},  we  expected  four  political
parties (UMP,  liberal party, UDF, and PS)  to play a key  role in the
network and  therefore we looked  for $Q=4$ clusters.  The $IL_{osbm}$
model selection  criterion now allows us to  estimate $Q^{*}$ directly
from the data, without any prior information. As we shall see, the PRG,
which  was  discarded  in  the  original study,  also  influences  the
topology of the network. 

\begin{figure}[htbp] \centering
  \setlength{\unitlength}{5mm}
  \includegraphics[width=10cm, height=10cm]{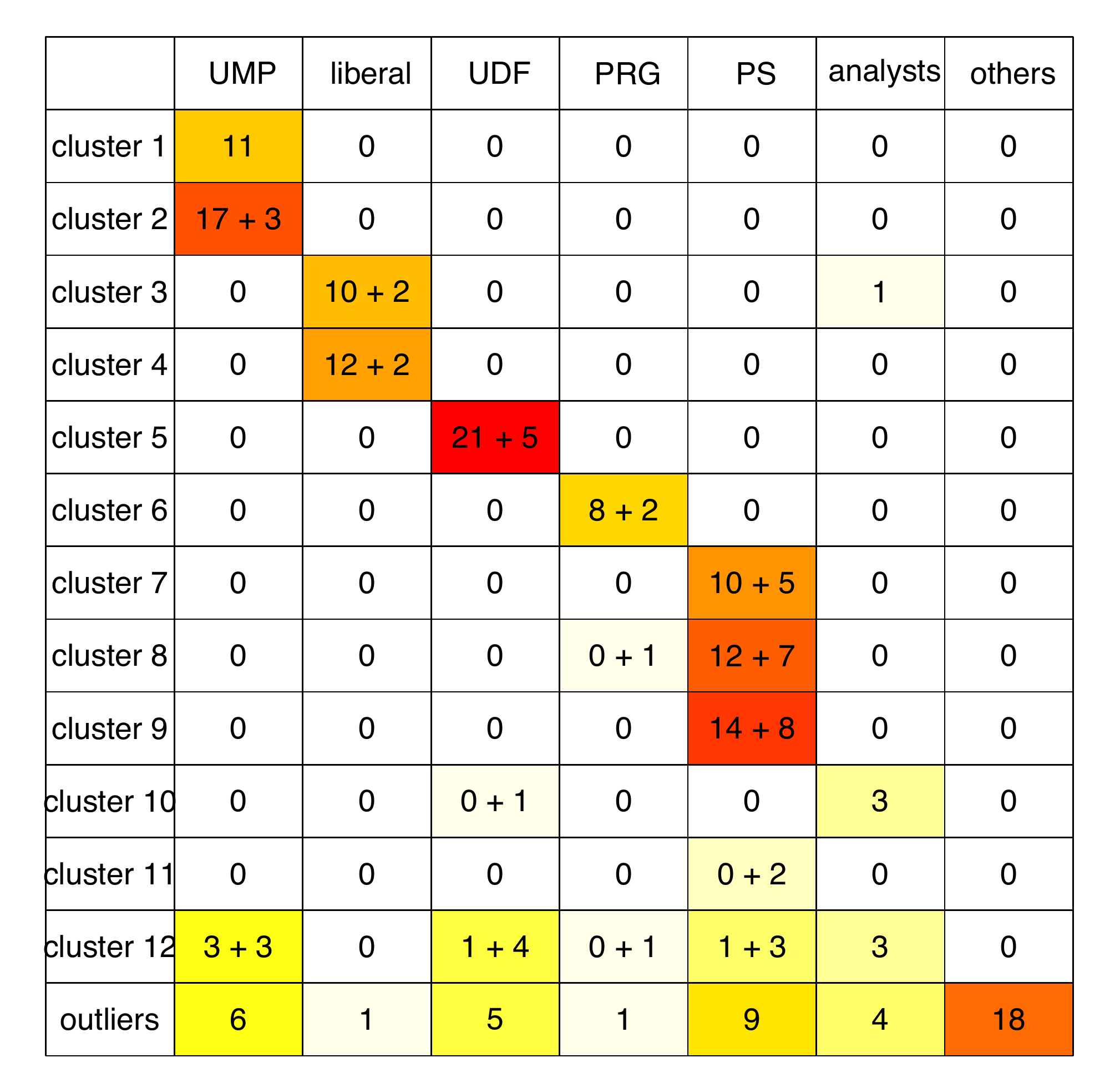}
  \caption{Classification of the blogs into $Q=12$ clusters using OSBM. The entry $(i, j)$ of the matrix describes the number of blogs associated to the $j$-th political party (column) and classified into cluster $i$ (row). Each entry distinguishes blogs which belong to a unique cluster from overlaps (single membership blogs $+$ overlaps). The last row corresponds to the null component.}
  \label{fig:resultsOSBM}
\end{figure}

We run the VBEM algorithm on the data sets for $Q \in \{1,\dots,
15\}$.  The $IL_{osbm}$  is computed  and such  procedure  is repeated
$100$  times, for  different initialization  of $\btau$.   Finally, we
select the best  learnt model for which the  model selection criterion
is  maximized. Thus,  we  find  $Q^{*}=12$ and  a  description of  the
corresponding clustering is given in Figure \ref{fig:resultsOSBM}.

First, we notice  that the first nine clusters  are highly homogeneous
and correspond  to well known  political parties. Thus, cluster  1 
contains  11  vertices which  are  all  associated  to UMP.  Moreover,
cluster  2  contains  20  vertices   all  associated  to  the  same
political party. Similarly, it follows that cluster 3 and 4 correspond
to the liberal party, cluster 5  to UDF, cluster 6 to PRG, and cluster
7,8,  9 to  PS. These  results are  relevant and  highlight  some 
interesting features  in the network. Indeed,  clustering the vertices
into  $Q=4$ clusters  as  in \cite{articlelatouche2011}  only gives  a
rough picture of the reality. In practice, the UMP, liberal party, and
PS  are organized  into several  clusters having  different connection
patterns. This  might indicate various political  affinities among the
political parties. The extreme case is for the PS which was split into
three  clusters.  Contrarily  to   the  original  study  where  PRG  was
discarded, most blogs associated to  PRG were classified into the same
cluster. This indicates that PRG plays  a role in shaping parts of the
network. Cluster  10 is  also homogeneous and  contains four  blogs among
which three correspond to blogs of political analysts. 

Cluster  11 is  of interest  because it  does not  contain  any single
membership blog. In other words,  its two blogs are both associated to
other clusters.  Thus, one  of them was  clustered in both  cluster 11
 and
cluster  9  (PS). Its  hostname  is ``www.parti-socialiste.fr''.  The
second was clustered in cluster 11 , cluster 7 (PS), and cluster 9
(PS).          The          corresponding         hostname          is
``annuaire.parti-socialiste.fr''. These two blogs are the most popular
blogs of PS, ``www.parti-socialiste.fr'' being the official website of the PS
itself, while  ``annuaire.parti-socialiste.fr'' lists all  the members
of PS. Interestingly, an extra  component was used for the clustering,
and these blogs  were not just found as  overlapping PS clusters, like
clusters 7 and 9. This can  be easily explained by the nature of these
blogs.  Indeed, contrarily to  the PS  blogs which  tend to  connect, as
other political parties, to blogs of their own party, these blogs have
extra connections to others. Blogs  of other political parties tend to
connect  to   them  simply   because  they  are   a  rich   source  of
information.  Finally, cluster 12 is an heterogeneous cluster, which
contains blogs of different political parties, from the left wing to the
right wing.  Interestingly, these blogs were classified  into the same
cluster due to their relation ties with the world of media.  In particular, we
point out that three of the blogs with single memberships are blogs of
political analysts. Moreover, all blogs from cluster 12 have been popular since
the French presidential election in 2007, most of them being mentioned
or referenced in newspapers. 

We  uncovered 23  overlaps in  the network  which are  described  in more
detail in  Table \ref{table:blogs}. As mentioned  previously, we found
that the  liberal party  and PS were  organized into  several clusters
corresponding to sub-groups having various political affinities. Therefore, it is of
no surprise  to find blogs  overlapping these clusters.  For instance,
two blogs associated  with the liberal party belong  to both cluster 3
(liberal)  and cluster  4 (liberal).  Furthermore,  PS is  made of  11
overlaps among which 10 are 2-membership and three-membership overlaps
between clusters 7, 8, 9, and 11 all corresponding to PS clusters. One
blog from  PRG overlaps cluster 6  (PRG) and cluster 8  (PS). This can
easily be understood since both PRG and PS are from the left wing and are known to
have  some relation  ties. Finally,  we emphasize  that  all political
parties, except the  liberal party, have overlaps with  cluster 12. We
recall that  this cluster contains  blogs with strong  connection with
the world of media.

In  the  original  study  in  \cite{articlelatouche2011},  with  $Q=4$
clusters, 59 blogs were
identified as  outliers and not classified.  With $Q^{*}=12$ clusters,
more  blogs  are  now  classified  and  only 44  blogs  are  found  as
outliers (null component). These blogs have weak connection profiles compared to all
the others.

\begin{table}[h]
  \centering
  \begin{tabular}{ c|c c c c c c c} 
  overlaps & UMP & liberal & UDF & PRG & PS & analysts & others \\
   \hline 
  clusters 2 (UMP)-12 (media) & $\mathbf{3}$ & 0 & 0 & 0 & 0 & 0 & 0\\
clusters 3 (liberal)-4 (liberal) & 0 & $\mathbf{2}$ & 0 & 0 & 0 & 0 & 0\\
clusters 5 (UDF)-12 (media) & 0 & 0 & $\mathbf{4}$ & 0 & 0 & 0 & 0\\
clusters 5 (UDF)-10 (media) & 0 & 0 & $\mathbf{1}$ & 0 & 0 & 0 & 0\\
clusters 6 (PRG)-8 (PS) & 0 & 0 & 0 & $\mathbf{1}$ & 0 & 0 & 0 \\
clusters 6 (PRG)-12 (media) & 0 & 0 & 0 & $\mathbf{1}$ & 0 & 0 & 0\\
clusters 7 (PS)-8 (PS) & 0 & 0 & 0 & 0 & $\mathbf{2}$ & 0 & 0 \\
clusters 8 (PS)-9 (PS) & 0 & 0 & 0 & 0 & $\mathbf{2}$ & 0 & 0 \\
clusters 7 (PS)-9 (PS) & 0 & 0 & 0 & 0 & $\mathbf{2}$ & 0 & 0 \\
clusters 9 (PS)-11 (PS) & 0 & 0 & 0 & 0 & $\mathbf{1}$ & 0 & 0 \\
clusters 7 (PS)-9 (PS)-11  (PS)& 0 & 0 & 0 & 0 &  $\mathbf{1}$ & 0 & 0
\\
clusters 8 (PS)-9 (PS)-12 (media) & 0 & 0 & 0 & 0 & $\mathbf{2}$ & 0 & 0
\\
clusters 8 (PS)-12 (media)& 0 & 0 & 0 & 0 &  $\mathbf{1}$ & 0 & 0
\\
   \end{tabular}
   \caption{Description of the $23$ overlaps found when clustering the
     blogs  into  $Q=12$ clusters  using  OSBM.  Non-zero entries  are
     indicated in bold. }
   \label{table:blogs}
\end{table}

\section{Conclusion}

In this  paper, we  proposed a Bayesian  rewriting of  the overlapping
stochastic  block model,  which led  us to  
 an estimation algorithm and an associated model selection criterion.
Introducing some conjugate prior distributions for the parameters of
OSBM, we  proposed a variational  Bayes EM algorithm, based  on global
and  local  variational  techniques.  The  algorithm can  be  used  to
approximate the  posterior distribution over the  model parameters and
latent  variables, given  the  observed data.  In  this framework,  we
derived a  model selection criterion, so called  $IL_{osbm}$, which is
based   on   a   non   asymptotic  approximation   of   the   marginal
log-likelihood.  Using simulated data  and a  real network,  we showed
that  $IL_{osbm}$ provides  a  relevant estimation  of  the number  of
overlapping  clusters. In  future work,   we are  interested in
exploring  parsimonious model  selection  in order  to choose  between
models where some of the network structure parameters
 $\bU$, $\bV$, $\bW$ and $W^{*}$ are set to zero or not.

\newpage
\section*{Appendix A: Appendix section}

\subsection{Lower Bound}\label{sec:lowBound1}

  Given a $N \times N$ positive real matrix $\bxi$, a lower bound of the first lower bound can be computed:
  \begin{equation*}
    \log p(\bX) \geq \LowerBound(q) \geq \LowerBound(q;\:\bxi),
  \end{equation*}
  where
  \begin{equation*}
    \LowerBound(q;\: \bxi) = \sum_{\bZ} \int \int \int q(\bZ, \balpha, \btW, \beta) \log\big(\frac{h(\bZ, \btW, \bxi)p(\bZ,\balpha,\btW,\beta)}{q(\bZ, \balpha, \btW, \beta)}\big) d\balpha d\btW d\beta,
  \end{equation*}
  and
  \begin{equation*}
     \log h(\bZ, \btW, \bxi) =  \sum_{i \neq j}^{N}\left\{(X_{ij}-\frac{1}{2})a_{\bZ_{i},\bZ_{j}} - \frac{\xi_{ij}}{2} + \log g(\xi_{ij}) - \lambda(\xi_{ij})(a_{\bZ_{i},\bZ_{j}}^{2} - \xi_{ij}^{2})\right\}.
  \end{equation*}

{\bf Proof:}
Let us start by showing that:
\begin{displaymath}
    \log p(\bX| \bZ, \btW) \geq \log h(\bZ, \btW, \bxi),
\end{displaymath}
  where $\bxi$ is an $N \times N$ positive real matrix.  We use the bound on the log-logistic function introduced by \cite{articlejaakkola2000}:
  \begin{equation} \label{eq:boundJaakkola}
    \log g(x) \geq \log g(\xi) + \frac{x-\xi}{2} - \lambda(\xi)(x^{2} - \xi^{2}),\forall (x, \xi) \in \mathbb{R}\times\mathbb{R}^{+},
  \end{equation}
where $\lambda(\xi) = (g(\xi)-1/2)/(2\xi)$.
 Note that (\ref{eq:boundJaakkola}) is an even function and therefore we can consider only positive values  of $x$ without loss of generality. 
  Since 
  \begin{displaymath}
    \log p(X_{ij}| \bZ_{i}, \bZ_{j}, \btW) = X_{ij}a_{\bZ_{i},\bZ_{j}} + \log g(-a_{\bZ_{i},\bZ_{j}}),
  \end{displaymath}
  then
  \begin{equation}
    \begin{aligned}
      \log p(X_{ij}| \bZ_{i}, \bZ_{j}, \btW) &\geq X_{ij}a_{\bZ_{i},\bZ_{j}} + \log g(\xi_{ij}) - \frac{a_{\bZ_{i},\bZ_{j}} + \xi_{ij}}{2} - \lambda(\xi_{ij})(a_{\bZ_{i},\bZ_{j}}^{2} - \xi_{ij}^{2}) \\
      &= (X_{ij}-\frac{1}{2})a_{\bZ_{i},\bZ_{j}} - \frac{\xi_{ij}}{2} + \log g(\xi_{ij}) - \lambda(\xi_{ij})(a_{\bZ_{i},\bZ_{j}}^{2} - \xi_{ij}^{2}).
    \end{aligned}
  \end{equation}
Following (\ref{eq:conDist}):
\begin{equation*}
  \log p(\bX| \bZ, \btW) = \sum_{i \neq j}^{N} \log p(\bX_{ij}| \bZ_{i}, \bZ_{j}, \btW).
\end{equation*}
Therefore
\begin{equation*}
  \log p(\bX| \bZ, \btW)\geq \log h(\bZ, \btW, \bxi).
\end{equation*}
We recall that the lower bound $\LowerBound(q)$ is given by:
\begin{equation*}
  \begin{aligned}
  \LowerBound(q) &= \sum_{\bZ}\int \int \int q(\bZ,\balpha,\btW,\beta) \log\left\{\frac{p(\bX,\bZ,\balpha,\btW,\beta)}{q(\bZ,\balpha,\btW,\beta)}\right\}d\balpha  d\btW d\beta\\
  &= \sum_{\bZ}\int \int \int q(\bZ,\balpha,\btW,\beta)\log p(\bX|\bZ,\btW)d\balpha d\btW d\beta  \\
  & \quad + \sum_{\bZ}\int \int \int q(\bZ,\balpha,\btW, \beta)\log\left\{\frac{p(\bZ,\balpha,\btW,\beta)}{q(\bZ,\balpha,\btW,\beta)}\right\}d\balpha d\btW d\beta \\
  & \geq \sum_{\bZ}\int \int \int q(\bZ,\balpha,\btW, \beta)\log h(\bZ,\btW,\bxi)d\balpha d\btW d\beta  \\
  & \quad + \sum_{\bZ} \int \int \int q(\bZ,\balpha,\btW,\beta)\log \left\{\frac{p(\bZ,\balpha,\btW,\beta)}{q(\bZ,\balpha,\btW,\beta)}\right\}d\balpha d\btW d\beta \\
  &= \sum_{\bZ} \int \int \int q(\bZ, \balpha, \btW, \beta) \log\left\{\frac{h(\bZ, \btW, \bxi)p(\bZ,\balpha,\btW,\beta)}{q(\bZ, \balpha, \btW,\beta)}\right\} d\balpha d\btW d\beta \\
  &= \LowerBound(q;\:\bxi).
  \end{aligned}
\end{equation*}
Finally
\begin{equation*}
  \log p(\bX) \geq \LowerBound(q) \geq \LowerBound(q;\:\bxi).
\end{equation*}

\subsection{Optimization of $q(\balpha)$}\label{sec:optAlpha}

The optimization of the lower bound with respect to $q(\balpha)$ produces a distribution with the same functional form as the prior $p(\balpha)$:
\begin{equation*}
  q(\balpha) = \prod_{q=1}^{Q} \Beta(\alpha_{q};\:\eta_{q}^{N},\zeta_{q}^{N}),
\end{equation*}
where
\begin{equation*}
  \eta_{q}^{N} = \eta_{q}^{0} + \sum_{i=1}^{N} \tau_{iq},
\end{equation*}
and
\begin{equation*}
  \zeta_{q}^{N} = \zeta_{q}^{0} + N - \sum_{i=1}^{N}\tau_{iq}.
\end{equation*}

{\bf Proof:}
According to variational Bayes, the optimal distribution $q(\balpha)$ is given by:
  \begin{equation} \label{eq:logqa}
    \begin{aligned}
      \log q(\balpha) &= \Esp_{\bZ, \btW, \beta}[\log\big(h(\bZ, \btW, \bxi)p(\bZ,\balpha,\btW,\beta)\big)] + \cst \\
      &= \Esp_{\bZ}[\log p(\bZ| \balpha)] + \log p(\balpha) + \cst \\
      &= \sum_{i=1}^{N}\sum_{q=1}^{Q}\left\{\tau_{iq}\log \alpha_{q} + (1-\tau_{iq})\log(1-\alpha_{q})\right\} + \sum_{q=1}^{Q}\left\{(\eta_{q}^{0}-1)\log\alpha_{q} + (\zeta_{q}^{0}-1)\log(1-\alpha_{q})\right\}\\
      &\quad + \cst \\
      &= \sum_{q=1}^{Q}\left\{(\eta_{q}^{0} + \sum_{i=1}^{N}\tau_{iq} - 1)\log \alpha_{q}+ (\zeta_{q}^{0} + N - \sum_{i=1}^{N}\tau_{iq} - 1)\log(1-\alpha_{q})\right\} + \cst.
    \end{aligned}
  \end{equation}
The functional form of (\ref{eq:logqa}) corresponds to the logarithm of a product of Beta distributions.

\subsection{Optimization of $q(\btW)$}\label{sec:optW}

The optimization of the lower bound with respect to $q(\btW)$ produces a distribution with the same functional form as the prior $p(\btW)$:
 \begin{equation*}
    q(\btW^{vec}) = \mathcal{N}(\btW^{\asvec};\: \btW_{N}^{\asvec}, \covmat_{N}), 
  \end{equation*}
  with
  \begin{equation*}
    \covmat_{N}^{-1} = \frac{a_{N}}{b_{N}}\bI + 2\sum_{i \neq j}^{N} \lambda(\xi_{ij})(\btE_{j}\otimes \btE_{i}),
  \end{equation*}
  and
  \begin{equation*}
    \btW_{N}^{\asvec} = \covmat_{N}\left\{\sum_{i \neq j}^{N} (X_{ij}-\frac{1}{2}) \bttau_{j}\otimes \bttau_{i}\right\}.
  \end{equation*}
  Each $(Q+1)\times(Q+1)$ probability matrix $\btE_{i}$ satisfies:
  \begin{equation*}
    \begin{aligned}
      \btE_{i} &= \Esp_{\bZ_{i}}[\btZ_{i}\btZ_{i}^{\intercal}]  \\
      &= \begin{pmatrix}
        \tau_{i1} & \tau_{i1}\tau_{i2} & \dots & \tau_{i1}\tau_{iQ} & \tau_{i1} \\
        \tau_{i2}\tau_{i1} & \tau_{i2} & \dots & \tau_{i2}\tau_{iQ} & \tau_{i2} \\ 
        \vdots & & & & \vdots \\
        \tau_{iQ}\tau_{i1} & \tau_{iQ}\tau_{i2} & \dots & \tau_{iQ} & \tau_{iQ} \\
        \tau_{i1} & \tau_{i2} & \dots & \tau_{iQ} & 1
      \end{pmatrix}.
    \end{aligned}
  \end{equation*}

{\bf Proof:}
According to variational Bayes, the optimal distribution $q(\btW)$ is given by:
  \begin{equation} \label{eq:decompW}
    \begin{aligned}
      \log q(\btW^{\asvec}) &= \Esp_{\bZ, \balpha,\beta}[\log \big(h(\bZ, \btW, \bxi)p(\bZ,\balpha,\btW,\beta)\big)] + \cst \\
      &= \Esp_{\bZ}[\log h(\bZ, \btW, \bxi)] + \Esp_{\beta}[\log p(\btW^{\asvec}|\beta)]  + \cst\\
      &= \sum_{i\neq j}^{N} \left\{(X_{ij}-\frac{1}{2})\Esp_{\bZ_{i},\bZ_{j}}[a_{\bZ_{i},\bZ_{j}}] - \lambda(\xi_{ij})\Esp_{\bZ_{i},\bZ_{j}}[a_{\bZ_{i},\bZ_{j}}^{2}] \right\} \\
      &\:\:\:\:\: - \frac{1}{2}\Esp_{\beta}[\beta](\btW^{\asvec})^{\intercal}\btW^{\asvec} + \cst.
    \end{aligned}
  \end{equation}
$\Esp_{\bZ_{i},\bZ_{j}}[a_{\bZ_{i},\bZ_{j}}]$ is given by:
\begin{equation} \label{eq:esp1W}
  \begin{aligned}
    \Esp_{\bZ_{i},\bZ_{j}}[a_{\bZ_{i},\bZ_{j}}] &= \Esp_{\bZ_{i},\bZ_{j}}[\btZ_{i}^{\intercal}\btW\btZ_{j}] \\
    &=\bttau_{i}^{\intercal}\btW\bttau_{j} \\
    &= (\bttau_{j} \otimes \bttau_{i})^{\intercal}\btW^{\asvec} \\
    &= (\btW^{\asvec})^{\intercal}(\bttau_{j}\otimes\bttau_{i}).
  \end{aligned}
\end{equation}
$\Esp_{\bZ_{i},\bZ_{j}}[a_{\bZ_{i},\bZ_{j}}^{2}]$ is given by:
\begin{equation} \label{eq:esp2W}
  \begin{aligned}
    \Esp_{\bZ_{i},\bZ_{j}}[a_{\bZ_{i},\bZ_{j}}^{2}] &= \Esp_{\bZ_{i},\bZ_{j}}[(\btZ_{i}^{\intercal}\btW\btZ_{j})^{2}]\\
    &= \Esp_{\bZ_{i},\bZ_{j}}[\big((\btZ_{j}\otimes\btZ_{i})^{\intercal}\btW^{\asvec}\big)^{2}] \\
    &= \Esp_{\bZ_{i},\bZ_{j}}[(\btZ_{j}\otimes\btZ_{i})^{\intercal}\btW^{\asvec}(\btZ_{j}\otimes\btZ_{i})^{\intercal}\btW^{\asvec}] \\
    &= \Esp_{\bZ_{i},\bZ_{j}}[(\btW^{\asvec})^{\intercal}(\btZ_{j}\otimes\btZ_{i})(\btZ_{j}\otimes\btZ_{i})^{\intercal}\btW^{\asvec}] \\
    &= \Esp_{\bZ_{i},\bZ_{j}}[(\btW^{\asvec})^{\intercal}\big((\btZ_{j}\btZ_{j}^{\intercal})\otimes(\btZ_{i}\btZ_{i}^{\intercal})\big)\btW^{\asvec}] \\
    &= (\btW^{\asvec})^{\intercal}\big(\btE_{j}\otimes\btE_{i}\big)\btW^{\asvec}.
  \end{aligned}
\end{equation}
$\Esp_{\beta}[\beta]$ is given by:
\begin{equation}\label{eq:esp3W}
  \Esp_{\beta}[\beta] = \frac{a_{N}}{b_{N}}.
\end{equation}

Using (\ref{eq:esp1W}), (\ref{eq:esp2W}) and (\ref{eq:esp3W}) in (\ref{eq:decompW}), we obtain:
\begin{equation}\label{eq:decompW2}
  \begin{aligned}
    \log q(\btW^{\asvec}) &= (\bW^{\asvec})^{\intercal}\left\{\sum_{i \neq j}^{N}(X_{ij}-\frac{1}{2})(\bttau_{j}\otimes\bttau_{i})\right\} \\
    &\:\:\:\:\:-\frac{1}{2}(\btW^{\asvec})^{\intercal}\left\{\frac{a_{N}}{b_{N}}\bI + 2\sum_{i \neq j}^{N}\lambda(\xi_{ij})\big(\btE_{j}\otimes\btE_{i}\big)\right\}\btW^{\asvec} + \cst.
  \end{aligned}
\end{equation}
The functional form of (\ref{eq:decompW2}) corresponds to the logarithm of a Gaussian distribution with mean $\btW^{\asvec}_{N}$ and covariance matrix $\covmat_{N}$.

\subsection{Optimization of $q(\beta)$}\label{sec:optBeta}
The optimization of the lower bound with respect to $q(\beta)$ produces a distribution with the same functional form as the prior $p(\beta)$:
  \begin{equation*}
    q(\beta) = \Gam(\beta;\:a_{N}, b_{N}), 
  \end{equation*}
where 
\begin{equation*}
  a_{N} = a_{0} + \frac{(Q+1)^{2}}{2}, 
\end{equation*}
and
\begin{equation*}
b_{N} = b_{0} + \frac{1}{2}\Trace(S_{N}) + \frac{1}{2}(\btW_{N}^{\asvec})^{\intercal}\btW_{N}^{\asvec}.
\end{equation*}

{\bf Proof:}
According to variational Bayes, the optimal distribution $q(\beta)$ is given by:
\begin{equation}\label{eq:decompBeta}
  \begin{aligned}
  \log q(\beta) &= \Esp_{\bZ,\balpha,\btW}[\log\big(h(\bZ, \btW, \bxi)p(\bZ,\balpha,\btW,\beta)\big)] + \cst \\
  &= \Esp_{\btW}[\log p(\btW|\beta)] + \log p(\beta) + \cst \\
  &=  -\frac{1}{2}\log |\frac{\bI}{\beta}| -\frac{\beta}{2}\Esp_{\btW}[(\btW^{\asvec})^{\intercal}\btW^{\asvec}] + (a_{0} - 1)\log \beta - b_{0}\beta + \cst \\
  &= \big(a_{0} + \frac{(Q+1)^{2}}{2} -1\big)\log \beta - \beta\Big(b_{0} + \frac{1}{2}\Trace(S_{N}) + \frac{1}{2}(\btW_{N}^{\asvec})^{\intercal}\btW_{N}^{\asvec}\Big) + \cst.
  \end{aligned}
\end{equation}
The functional form of (\ref{eq:decompBeta}) corresponds to the logarithm of a Gamma distribution.

\subsection{Optimization of $q(Z_{iq})$}\label{sec:optZiq}
The optimization of the lower bound with respect to $q(Z_{iq})$ produces a distribution with the same functional form as the prior $p(Z_{iq}|\balpha)$:
 \begin{equation*}
    q(Z_{iq}) = \Bernoulli(Z_{iq};\:\tau_{iq}),
  \end{equation*}

where
\begin{equation*}
  \begin{aligned}
\tau_{iq} &= g\bigg\{\psi(\eta_{q}^{N})-\psi(\zeta_{q}^{N}) + \sum_{j\neq i}^{N}(X_{ij}-\frac{1}{2})\bttau_{j}^{\intercal}(\btW_{N}^{\intercal})_{\cdot q} + \sum_{j \neq i}^{N}(X_{ji}-\frac{1}{2})\bttau_{j}^{\intercal}(\btW_{N})_{\cdot q}  \\
    &\:\:\:\:\: -\Trace\Big(\big(\bSigma_{qq}^{'} + 2\sum_{l\neq q}^{Q+1}\ttau_{il}\bSigma_{ql}^{'}\big)\big(\sum_{j\neq i}^{N}\lambda(\xi_{ij})\btE_{j}\big) + \big(\bSigma_{qq} + 2\sum_{l\neq q}^{Q+1} \ttau_{il}\bSigma_{ql}\big)\big(\sum_{j\neq i}^{N}\lambda(\xi_{ji})\btE_{j}\big)\Big)\bigg\},
    \end{aligned}
\end{equation*}
and $\bSigma_{ql}=\Esp_{\btW_{q},\btW_{l}}[\btW_{\cdot q} \btW_{\cdot l}^{\intercal}]$, $\bSigma_{ql}^{'} = \Esp_{\btW_{q\cdot},\btW_{l\cdot}}[\btW_{q\cdot}^{\intercal}\btW_{l\cdot}]$.

{\bf Proof:}
According to variational Bayes, the optimal distribution $q(Z_{iq})$ is given by:
  \begin{equation*}
    \log q(Z_{bc}) = \Esp_{\bZ^{\backslash bc},\balpha, \btW, \beta}[\log\big(h(\bZ, \btW, \bxi)p(\bZ,\balpha,\btW,\beta)\big)] + \cst,
  \end{equation*}
  where $\bZ^{\backslash bc}$ is the set of all class memberships except $Z_{bc}$.
  \begin{equation*}
    \log q(Z_{bc}) = \Esp_{\bZ^{\backslash bc}, \btW}[\log h(\bZ, \btW, \bxi)] + \Esp_{\bZ^{\backslash bc},\balpha}[\log p(\bZ|\balpha)] + \cst.
  \end{equation*}
$\Esp_{\bZ^{\backslash bc},\balpha}[\log p(\bZ|\balpha)]$ is given by:
\begin{equation*}
  \begin{aligned}
    \Esp_{\bZ^{\backslash bc},\balpha}[\log p(\bZ|\balpha)] &= Z_{bc}\Esp_{\alpha_{c}}[\log \alpha_{c}] + (1 - Z_{bc})\Esp_{\alpha_{c}}[\log(1-\alpha_{c})] + \cst \\
    &= Z_{bc}\big(\psi(\eta_{c}^{N})-\psi(\eta_{c}^{N}+\zeta_{c}^{N})\big) + (1-Z_{bc})\big(\psi(\zeta_{c}^{N})-\psi(\eta_{c}^{N} + \zeta_{c}^{N})\big) + \cst \\
    &= Z_{bc}\big(\psi(\eta_{c}^{N}) - \psi(\zeta_{c}^{N})\big) + \cst,
  \end{aligned}
\end{equation*}
where $\psi(\cdot)$ is the digamma function (the logarithmic derivative of the gamma function $\Gamma(\cdot)$ which appears in the normalizing constants of the Beta distributions).

\begin{equation*}
  \begin{aligned}
    \Esp_{\bZ^{\backslash bc}, \btW}[\log h(\bZ, \btW, \bxi)]&=\sum_{i \neq j}^{N}\left\{(X_{ij}-\frac{1}{2})\Esp_{\bZ^{\backslash bc},\btW}[a_{\bZ_{i},\bZ_{j}}] - \lambda(\xi_{ij})\Esp_{\bZ^{\backslash bc},\btW}[a_{\bZ_{i},\bZ_{j}}^{2}]\right\} + \cst \\
    &=\sum_{j\neq b}^{N} \left\{(X_{bj}-\frac{1}{2})\Esp_{\bZ_{b}^{\backslash c},\bZ_{j},\btW}[a_{\bZ_{b},\bZ_{j}}] - \lambda(\xi_{bj})\Esp_{\bZ_{b}^{\backslash c},\bZ_{j},\btW}[a_{\bZ_{b},\bZ_{j}}^{2}]\right\} \\
&\:\:\:\:\:+\sum_{i\neq b}^{N} \left\{(X_{ib}-\frac{1}{2})\Esp_{\bZ_{b}^{\backslash c},\bZ_{i},\btW}[a_{\bZ_{i},\bZ_{b}}] - \lambda(\xi_{ib})\Esp_{\bZ_{b}^{\backslash c},\bZ_{i},\btW}[a_{\bZ_{i},\bZ_{b}}^{2}]\right\} + \cst\\
&=\sum_{j \neq b}^{N}\Big\{(X_{bj}-\frac{1}{2})\Esp_{\bZ_{b}^{\backslash c},\bZ_{j},\btW}[a_{\bZ_{b},\bZ_{j}}] + (X_{jb}-\frac{1}{2})\Esp_{\bZ_{b}^{\backslash c},\bZ_{j},\btW}[a_{\bZ_{j},\bZ_{b}}] \\
&\:\:\:\:\: -\lambda(\xi_{bj})\Esp_{\bZ_{b}^{\backslash c},\bZ_{j},\btW}[a_{\bZ_{b},\bZ_{j}}^{2}] - \lambda(\xi_{jb})\Esp_{\bZ_{b}^{\backslash c},\bZ_{j},\btW}[a_{\bZ_{j},\bZ_{b}}^{2}]\Big\} + \cst.
    \end{aligned}
\end{equation*}

\begin{equation*} 
  \begin{aligned}
    \Esp_{\bZ_{b}^{\backslash c},\bZ_{j},\btW}[a_{\bZ_{b},\bZ_{j}}] &= \Esp_{\bZ_{b}^{\backslash c},\bZ_{j},\btW}[\sum_{q,l}^{Q+1} \tZ_{bq}\tW_{ ql}\tZ_{jl}] \\
    &= Z_{bc}\sum_{l=1}^{Q+1}\Esp_{\tW_{cl}}[\tW_{cl}]\ttau_{jl} + \cst \\
    &= Z_{bc}\bttau_{j}^{\intercal}(\btW_{N}^{\intercal})_{\cdot c} + \cst.
  \end{aligned}
\end{equation*}
\begin{equation*}
  \begin{aligned}
    \Esp_{\bZ_{b}^{\backslash c},\bZ_{j},\btW}[a_{\bZ_{j},\bZ_{b}}] &= \Esp_{\bZ_{b}^{\backslash c},\bZ_{j},\btW}[\sum_{q,l}^{Q+1}\tZ_{jq}\tW_{ql}\tZ_{bl}] \\
    &= Z_{bc}\sum_{l=1}^{Q+1}\Esp_{\tW_{lc}}[\tW_{lc}]\ttau_{jl} + \cst \\
    &= Z_{bc}\bttau_{j}^{\intercal}(\btW_{N})_{\cdot c} + \cst.
  \end{aligned}
\end{equation*}

\begin{equation*}
  \begin{aligned}
    \Esp_{\bZ_{b}^{\backslash c},\bZ_{j},\btW}[a_{\bZ_{j},\bZ_{b}}^{2}] &= \Esp_{\bZ_{b}^{\backslash c},\bZ_{j},\btW}[\big(\sum_{q,l}^{Q+1}\tZ_{jq}\tW_{ql}\tZ_{bl}\big)\big(\sum_{q,l}^{Q+1}\tZ_{jq}\tW_{ql}\tZ_{bl}\big)] \\
    &= \Esp_{\bZ_{b}^{\backslash c},\bZ_{j},\btW}[\sum_{q,q^{'},l,l^{'}}^{Q+1}\tZ_{bl}\tZ_{bl^{'}}\tZ_{jq}\tW_{ql}\tW_{q^{'}l^{'}}\tZ_{jq^{'}}] \\
    &= \Esp_{\bZ_{b}^{\backslash c},\bZ_{j},\btW}[Z_{bc}\sum_{q,q^{'}}^{Q+1}\tZ_{jq}\tW_{qc}\tW_{q^{'}c}\tZ_{jq^{'}} + 2Z_{bc}\sum_{q, q^{'},l\neq c}^{Q+1}\tZ_{bl}\tZ_{jq}\tW_{qc}\tW_{q^{'}l}\tZ_{jq^{'}}] + \cst \\
    &=Z_{bc}\left\{\Esp_{\bZ_{j},\btW_{\cdot c}}[\btW_{\cdot c}^{\intercal}\btZ_{j}\btZ_{j}^{\intercal}\btW_{\cdot c}] + 2\sum_{l\neq c}^{Q+1} \ttau_{bl}\Esp_{\bZ_{j},\btW_{\cdot c},\btW_{\cdot l}}[\btW_{\cdot c}^{\intercal}\btZ_{j}\btZ_{j}^{\intercal}\btW_{\cdot l}]\right\} + \cst \\
    &=Z_{bc}\left\{\Esp_{\btW_{\cdot,c}}[\btW_{\cdot c}^{\intercal}\btE_{j}\btW_{\cdot c}] + 2\sum_{l\neq c}^{Q+1} \ttau_{bl}\Esp_{\btW_{\cdot c},\btW_{\cdot l}}[\btW_{\cdot c}^{\intercal}\btE_{j}\btW_{\cdot l}]\right\} + \cst \\
    &=Z_{bc}\left\{\Esp_{\btW_{\cdot c}}[(\btW_{\cdot c}\otimes \btW_{\cdot c})^{\intercal}]\btE_{j}^{\asvec} + 2\sum_{l \neq c}^{Q+1}\ttau_{bl}\Esp_{\btW_{\cdot c},\btW_{\cdot l}}[(\btW_{\cdot l} \otimes \btW_{\cdot c})^{\intercal}]\btE_{j}^{\asvec}\right\} + \cst \\
    &= Z_{bc}\left\{\Esp_{\btW_{\cdot c}}[((\btW_{\cdot c}\btW_{\cdot c}^{\intercal})^{\asvec})^{\intercal}]\btE_{j}^{\asvec} + 2\sum_{l\neq c}^{Q+1}\ttau_{bl}\Esp_{\btW_{\cdot c},\btW_{\cdot l}}[((\btW_{\cdot c}\btW_{\cdot l}^{\intercal})^{\asvec})^{\intercal}]\btE_{j}^{\asvec}\right\} + \cst \\
    &= Z_{bc}\left\{(\bSigma_{cc}^{\asvec})^{\intercal}\btE_{j}^{\asvec} + 2\sum_{l\neq c}^{Q+1} \ttau_{bl}(\bSigma_{cl}^{\asvec})^{\intercal}\btE_{j}^{\asvec}\right\} + \cst \\
    &=Z_{bc}\Trace\Big(\big(\bSigma_{cc} + 2\sum_{l\neq c}^{Q+1}\ttau_{bl}\bSigma_{cl}\big)\btE_{j}\Big) + \cst, 
    \end{aligned}
    \end{equation*}
where $\bSigma_{ql}=\Esp_{\btW_{q},\btW_{l}}[\btW_{\cdot q} \btW_{\cdot l}^{\intercal}]$.
Similarly, we have:
\begin{equation*}
  \Esp_{\bZ_{b}^{\backslash c},\bZ_{j},\btW}[a_{\bZ_{b},\bZ_{j}}^{2}] = Z_{bc}\Trace\Big(\big(\bSigma_{cc}^{'} + 2\sum_{l\neq c}^{Q+1}\ttau_{bl}\bSigma_{cl}^{'}\big)\btE_{j}\Big) + \cst,
\end{equation*}
where $\bSigma_{ql}^{'} = \Esp_{\btW_{q\cdot},\btW_{l\cdot}}[\btW_{q\cdot}^{\intercal}\btW_{l\cdot}]$.
Finally, we obtain:
\begin{equation} \label{eq:logqZbc}
  \begin{aligned}
    \log q(Z_{bc}) &= Z_{bc}\bigg\{\psi(\eta_{c}^{N})-\psi(\zeta_{c}^{N}) + \sum_{j\neq b}^{N}(X_{bj}-\frac{1}{2})\bttau_{j}^{\intercal}(\btW_{N}^{\intercal})_{\cdot c} + \sum_{j \neq b}^{N}(X_{jb}-\frac{1}{2})\bttau_{j}^{\intercal}(\btW_{N})_{\cdot c}  \\
    &\:\:\:\:\: -\Trace\Big(\big(\bSigma_{cc}^{'} + 2\sum_{l\neq c}^{Q+1}\ttau_{bl}\bSigma_{cl}^{'}\big)\big(\sum_{j\neq b}^{N}\lambda(\xi_{bj})\btE_{j}\big) + \big(\bSigma_{cc} + 2\sum_{l\neq c}^{Q+1} \ttau_{bl}\bSigma_{cl}\big)\big(\sum_{j\neq b}^{N}\lambda(\xi_{jb})\btE_{j}\big)\Big)\bigg\}  \\
    &\quad + \cst.
  \end{aligned}
\end{equation}
The functional form of (\ref{eq:logqZbc}) corresponds to the logarithm of a Bernoulli distribution with parameter $\tau_{bc}$. Indeed:
\begin{equation*}
  \begin{aligned}
    \log \Bernoulli(Z_{bc};\:\tau_{bc}) &= Z_{bc}\log\tau_{bc} + (1-Z_{bc})\log (1-\tau_{bc})   \\
    &=    Z_{bc}\log(\frac{\tau_{bc}}{1-\tau_{bc}}) + \cst. \\
  \end{aligned}
\end{equation*}
If we denote $p=\log\big(\tau_{bc}/(1-\tau_{bc})\big)$, then $\tau_{bc}=g(p)$.

\subsection{Optimization of $\bxi$}\label{sec:optXi}

Setting the partial derivative of the lower bound with respect to $\xi_{ij}$, to zero, leads to an estimate $\hat{\xi_{ij}}$ of $\xi_{ij}$:
  \begin{equation*}
    \hat{\xi}_{ij}= \sqrt{\Trace\Big(\big(\covmat_{N} + \btW_{N}^{\asvec}(\btW_{N}^{\asvec})^{\intercal}\big)(\btE_{j}\otimes \btE_{i})\Big)}.
  \end{equation*}

{\bf Proof:}
The partial derivative of the lower bound with respect to $\xi_{ij}$ is given by:
  \begin{equation*}
    \frac{\partial \LowerBound}{\partial \xi_{ij}}(q;\:\bxi) = -\frac{1}{2} + g(-\xi_{ij}) - \lambda^{'}(\xi_{ij})\big(\Esp_{\bZ_{i},\bZ_{j},\btW}[a_{\bZ_{i},\bZ_{j}}^{2}] - \xi_{ij}^{2}\big) + 2\xi_{ij}\lambda(\xi_{ij}).
  \end{equation*}
According to (\ref{eq:esp2W}), 
\begin{equation*}
  \Esp_{\bZ_{i},\bZ_{j}}[a_{\bZ_{i},\bZ_{j}}^{2}] = (\btW^{\asvec})^{\intercal}\big(\btE_{j}\otimes \btE_{i}\big)\btW^{\asvec},
\end{equation*}
therefore
\begin{equation} \label{eq:espDist}
  \begin{aligned}
  \Esp_{\bZ_{i},\bZ_{j},\btW}[a_{\bZ_{i},\bZ_{j}}^{2}] &= \Esp_{\btW}[(\btW^{\asvec})^{\intercal}(\btE_{j}\otimes \btE_{i})\btW^{\asvec}] \\
  &= \Esp_{\btW}\Big[[\Trace\Big(\btW^{\asvec}(\btW^{\asvec})^{\intercal}(\btE_{j}\otimes \btE_{i})\Big)\Big] \\
  &= \Trace\Big(\Esp_{\btW} \big[\btW^{\asvec}(\btW^{\asvec})^{\intercal}\big](\btE_{j}\otimes \btE_{i})\Big) \\
  &= \Trace\Big(\big(\covmat_{N} + \btW_{N}^{\asvec}(\btW_{N}^{\asvec})^{\intercal}\big)(\btE_{j}\otimes \btE_{i})\Big).
  \end{aligned}
\end{equation}
Moreover $(\log g)^{'}(\xi_{ij})= g(-\xi_{ij})$ and $g(\xi_{j}) + g(-\xi_{ij})=1$. We obtain:
\begin{equation*} \label{eq:optiXi}
  \frac{\partial \LowerBound}{\partial \xi_{ij}}(q;\:\bxi) = - \lambda^{'}(\xi_{ij})\left\{\Trace\Big(\big(\covmat_{N} + \btW_{N}^{\asvec}(\btW_{N}^{\asvec})^{\intercal}\big)(\btE_{j}\otimes \btE_{i})\Big) - \xi_{ij}^{2}\right\}.
\end{equation*}
Finally,  $\lambda(\xi_{ij})$ is a strictly decreasing function for positive values of $\xi_{ij}$. Thus,  $\lambda^{'}(\xi_{ij}) \neq 0$ and if we set the derivative of (\ref{eq:optiXi}) to zero, it leads to:
\begin{equation*}
\xi_{ij}^{2} = \Trace\Big(\big(\covmat_{N} + \btW_{N}^{\asvec}(\btW_{N}^{\asvec})^{\intercal}\big)(\btE_{j}\otimes \btE_{i})\Big).
\end{equation*}

\subsection{Lower bound}\label{sec:lowProof}

After the variational Bayes M-step, most of the terms in the lower bound vanish:
\begin{multline}
  \LowerBound(q;\:\bxi) = \sum_{i\neq j}^{N} \left\{\log g(\xi_{ij}) - \frac{\xi_{ij}}{2} + \lambda(\xi_{ij})\xi_{ij}^{2}\right\} + \sum_{q=1}^{Q}\log \bigg\{\frac{\Gamma(\eta_{q}^{0}+\zeta_{q}^{0})\Gamma(\eta_{q}^{N})\Gamma(\zeta_{q}^{N})}{\Gamma(\eta_{q}^{0})\Gamma(\zeta_{q}^{0})\Gamma(\eta_{q}^{N}+\zeta_{q}^{N})}\bigg\} + \log \frac{\Gamma(a_{N})}{\Gamma(a_{0})} + a_{0}\log b_{0} \\
+ a_{N}(1 - \frac{b_{0}}{b_{N}} - \log b_{N}) + \frac{1}{2}(\btW_{N}^{\asvec})^{\intercal}\covmat_{N}^{-1}\btW_{N}^{\intercal} + \frac{1}{2}\log |\covmat_{N}| - \sum_{i=1}^{N}\sum_{q=1}^{Q}\left\{\tau_{iq}\log \tau_{iq} + (1-\tau_{iq})\log(1-\tau_{iq})\right\}.
\end{multline}

{\bf Proof:}
  \begin{equation} \label{eq:lowerBoundDecomp}
    \begin{aligned}
      \LowerBound(q;\:\bxi) &= \sum_{\bZ}\int \int \int q(\bZ,\balpha,\btW,\beta)\log \big(\frac{h(\bZ, \btW, \bxi)p(\bZ,\balpha,\btW,\beta)}{q(\bZ,\balpha,\btW,\beta)}d\balpha d\btW d\beta\\
      &= \Esp_{\bZ,\btW}[\log h(\bZ,\btW,\bxi)] + \Esp_{\bZ,\balpha}[\log p(\bZ|\balpha)] + \Esp_{\balpha}[\log p(\balpha)] + \Esp_{\btW,\beta}[\log p(\btW|\beta)] + \Esp_{\beta}[\log p(\beta)] \\
      &\:\:\:\:\: - \Esp_{\bZ}[\log q(\bZ)] - \Esp_{\balpha}[\log q(\balpha)] - \Esp_{\btW}[\log q(\btW)] - \Esp_{\beta}[\log q(\beta)]\\
      &=\sum_{i\neq j}^{N}\bigg\{(X_{ij}-\frac{1}{2})\Esp_{\bZ_{i},\bZ_{j},\btW}[a_{\bZ_{i},\bZ_{j}}] - \frac{\xi_{ij}}{2} + \log g(\xi_{ij}) - \lambda(\xi_{ij})\big(\Esp_{\bZ_{i},\bZ_{j},\btW}[a_{\bZ_{i},\bZ_{j}}^{2}] - \xi_{ij}^{2}\big)\bigg\} \\
      &\:\:\:\:\: + \sum_{i=1}^{N}\sum_{q=1}^{Q}\bigg\{\tau_{iq}\big(\psi(\eta_{q}^{N}) - \psi(\eta_{q}^{N} + \zeta_{q}^{N})\big) + (1-\tau_{iq})\big(\psi(\zeta_{q}^{N})-\psi(\eta_{q}^{N}+\zeta_{q}^{N})\big)\bigg\} \\
      &\:\:\:\:\: + \sum_{q=1}^{Q}\bigg\{ \log (\frac{\Gamma(\eta_{q}^{0}+\zeta_{q}^{0})}{\Gamma(\eta_{q}^{0})\Gamma(\zeta_{q}^{0})}) + (\eta_{q}^{0}-1)\big(\psi(\eta_{q}^{N}) - \psi(\eta_{q}^{N} + \zeta_{q}^{N})\big) \\
      &\:\:\:\:\:    + (\zeta_{q}^{0}-1)\big(\psi(\zeta_{q}^{N})-\psi(\eta_{q}^{N}+\zeta_{q}^{N})\big)\bigg\} + \Esp_{\btW,\beta}[\log p(\btW|\beta)] - \log \Gamma(a_{0}) + a_{0}\log b_{0} \\ 
&\:\:\:\:\: + (a_{0}-1)\big(\psi(a_{N}) - \log b_{N}\big) - b_{0}\frac{a_{N}}{b_{N}} - \sum_{i=1}^{N}\sum_{q=1}^{Q}\bigg\{\tau_{iq}\log \tau_{iq} + (1-\tau_{iq})\log(1-\tau_{iq})\bigg\} \\
      &\:\:\:\:\: - \sum_{q=1}^{Q}\bigg\{\log(\frac{\Gamma(\eta_{q}^{N}+\zeta_{q}^{N})}{\Gamma(\eta_{q}^{N})\Gamma(\zeta_{q}^{N})}) + (\eta_{q}^{N}-1)\big(\psi(\eta_{q}^{N})-\psi(\eta_{q}^{N}+\zeta_{q}^{N})\big) \\
      &\:\:\:\:\:  + (\zeta_{q}^{N}-1)\big(\psi(\zeta_{q}^{N})-\psi(\eta_{q}^{N}+\zeta_{q}^{N})\big)\bigg\}- \Esp_{\btW}[\log q(\btW)] + \log \Gamma(a_{N}) - a_{N}\log b_{N} \\
      &\:\:\:\:\: - (a_{N}-1)\big(\psi(a_{N}) - \log b_{N}\big) + b_{N}\frac{a_{N}}{b_{N}}.
    \end{aligned}
  \end{equation}
$\Esp_{\bZ_{i},\bZ_{j},\btW}[a_{\bZ_{i},\bZ_{j}}]$ is given by:
\begin{equation} \label{eq:espInteract}
  \begin{aligned}
    \Esp_{\bZ_{i},\bZ_{j},\btW}[a_{\bZ_{i},\bZ_{j}}] &= \Esp_{\bZ_{i},\bZ_{j},\btW}[\btZ_{i}^{\intercal}\btW\btZ_{j}] \\
     &=\Esp_{\btW}[\bttau_{i}^{\intercal}\btW\bttau_{j}] \\
     &= \Esp_{\btW}[(\bttau_{j}\otimes\bttau_{i})^{\intercal}\btW^{\asvec}] \\
     &= \Esp_{\btW}[(\btW^{\asvec})^{\intercal}(\bttau_{j}\otimes \bttau_{i})] \\
     &= (\btW_{N}^{\asvec})^{\intercal}(\bttau_{j}\otimes \bttau_{i}).
  \end{aligned}
\end{equation}
$\Esp_{\bZ_{i},\bZ_{j},\btW}[a_{\bZ_{i},\bZ_{j}}^{2}]$ is given by (\ref{eq:espDist}) 

$\Esp_{\btW}[\log p(\btW|\beta)]$ is given by:
\begin{equation} \label{eq:espLogpw}
  \begin{aligned}
    \Esp_{\btW}[\log p(\btW|\beta)] &= -\frac{(Q+1)^{2}}{2}\log 2\pi - \frac{1}{2}\Esp_{\beta}[\log |\frac{\bI}{\beta}|] - \frac{1}{2}\Esp_{\beta}[\beta]\Esp_{\btW}[(\btW^{\asvec})^{\intercal}\btW^{\asvec}] \\
    &= -\frac{(Q+1)^{2}}{2}\log 2\pi + \frac{(Q+1)^{2}}{2}\Esp_{\beta}[\log \beta] - \frac{a_{N}}{2b_{N}}\Trace\Big(\covmat_{N} + \btW_{N}^{\asvec}(\btW_{N}^{\asvec})^{\intercal}\Big) \\ 
    &= -\frac{(Q+1)^{2}}{2}\log 2\pi + \frac{(Q+1)^{2}}{2}\big(\psi(a_{N}) - \log b_{N}\big) - \frac{a_{N}}{2b_{N}}\Trace\Big(\covmat_{N} + \btW_{N}^{\asvec}(\btW_{N}^{\asvec})^{\intercal}\Big).
  \end{aligned}
\end{equation}
Similarly, we have:
\begin{equation} \label{eq:espLogqw}
  \begin{aligned}
    \Esp_{\btW}[\log q(\btW)] &= -\frac{(Q+1)^{2}}{2}\log 2\pi - \frac{1}{2}\log|\covmat_{N}| - \frac{1}{2}\Esp_{\btW}[(\btW^{\asvec})^{\intercal}\covmat_{N}^{-1}\btW^{\asvec}] + \Esp_{\btW}[(\btW^{\asvec})^{\intercal}\covmat_{N}^{-1}\btW_{N}^{\asvec}] \\
    &\:\:\:\:\: - \frac{1}{2}(\btW_{N}^{\asvec})^{\intercal}\covmat_{N}^{-1}\btW_{N}^{\asvec} \\ 
    &= -\frac{(Q+1)^{2}}{2}\log 2\pi - \frac{1}{2}\log|\covmat_{N}| - \frac{1}{2}\Esp_{\btW}\Big[\Trace\Big(\btW^{\asvec}(\btW^{\asvec})^{\intercal}\covmat_{N}^{-1}\Big)\Big] + (\btW_{N}^{\asvec})^{\intercal}\covmat_{N}^{-1}\btW_{N}^{\asvec} \\
  &\:\:\:\:\: - \frac{1}{2}(\btW_{N}^{\asvec})^{\intercal}\covmat_{N}^{-1}\btW_{N}^{\asvec} \\
    &= -\frac{(Q+1)^{2}}{2}\log 2\pi - \frac{1}{2}\log|\covmat_{N}| - \frac{1}{2}\Trace\Big(\big(\covmat_{N} + \btW_{N}^{\asvec}(\btW_{N}^{\asvec})^{\intercal}\big)\covmat_{N}^{-1}\Big) + (\btW_{N}^{\asvec})^{\intercal}\covmat_{N}^{-1}\btW_{N}^{\asvec} \\
  &\:\:\:\:\: - \frac{1}{2}(\btW_{N}^{\asvec})^{\intercal}\covmat_{N}^{-1}\btW_{N}^{\asvec} 
  \end{aligned}
\end{equation}
    
 After rearranging the terms in (\ref{eq:lowerBoundDecomp}) and using (\ref{eq:espDist}), (\ref{eq:espInteract}), (\ref{eq:espLogpw}), as well as  (\ref{eq:espLogqw}), we obtain:
\begin{multline}
    \LowerBound(q;\:\bxi) = \sum_{i\neq j}^{N} \left\{\log g(\xi_{ij}) - \frac{\xi_{ij}}{2} + \lambda(\xi_{ij})\xi_{ij}^{2}\right\} + \sum_{q=1}^{Q}\log \bigg\{\frac{\Gamma(\eta_{q}^{0}+\zeta_{q}^{0})\Gamma(\eta_{q}^{N})\Gamma(\zeta_{q}^{N})}{\Gamma(\eta_{q}^{0})\Gamma(\zeta_{q}^{0})\Gamma(\eta_{q}^{N}+\zeta_{q}^{N})}\bigg\} + \log \frac{\Gamma(a_{N})}{\Gamma(a_{0})} + a_{0}\log b_{0} \\
+ a_{N}(1 - \frac{b_{0}}{b_{N}} - \log b_{N}) + \frac{1}{2}(\btW_{N}^{\asvec})^{\intercal}\covmat_{N}^{-1}\btW_{N}^{\intercal} + \frac{1}{2}\log |\covmat_{N}| - \sum_{i=1}^{N}\sum_{q=1}^{Q}\left\{\tau_{iq}\log \tau_{iq} + (1-\tau_{iq})\log(1-\tau_{iq})\right\} \\+ (a_{0} + \frac{(Q+1)^{2}}{2} - a_{N})\big(\psi(a_{N}) - \log b_{N}\big) \\ +\sum_{q=1}^{Q}\bigg\{\big(\eta_{q}^{0}+\sum_{i\neq j}^{N}\tau_{iq}-\eta_{q}^{N}\big)\big(\psi(\eta_{q}^{N})-\psi(\eta_{q}^{N}+\zeta_{q}^{N})\big) + \big(\zeta_{q}^{0} + N - \sum_{i=1}^{N}\tau_{iq}-\zeta_{q}^{N}\big)\big(\psi(\zeta_{q}^{N})-\psi(\eta_{q}^{N}+\zeta_{q}^{N})\big)\bigg\} \\
-\frac{1}{2}\Trace\Big(\big(\covmat_{N}+\btW_{N}^{\asvec}(\btW_{N}^{\asvec})^{\intercal}\big)\big(\frac{a_{N}}{b_{N}}\bI + 2\sum_{i\neq j}^{N} \lambda(\xi_{ij})(\btE_{j}\otimes \btE_{i})-\covmat_{N}^{-1}\big)\Big) \\
+                       (\btW_{N}^{\asvec})^{\intercal}\Big(\sum_{i\neq
  j}^{N}(X_{ij}-\frac{1}{2})(\bttau_{j}\otimes\bttau_{i})-\covmat_{N}^{-1}\btW_{N}^{\asvec}\Big)
. \\
\end{multline}
After the variational M step (optimization of $q(\btW)$), many terms vanish.

\bibliographystyle{plainnat}      
\bibliography{biblio}   

\begin{thebibliography}{43}
\providecommand{\natexlab}[1]{#1}
\providecommand{\url}[1]{\texttt{#1}}
\expandafter\ifx\csname urlstyle\endcsname\relax
  \providecommand{\doi}[1]{doi: #1}\else
  \providecommand{\doi}{doi: \begingroup \urlstyle{rm}\Url}\fi

\bibitem[Airoldi et~al.(2006)Airoldi, Blei, Xing, and
  Fienberg]{proceedingsairoldi2006}
E.~Airoldi, D.~Blei, E.~Xing, and S.~Fienberg.
\newblock Mixed membership stochastic block models for relational data with
  application to protein-protein interactions.
\newblock In \emph{Proceedings of the International Biometrics Society Annual
  Meeting}, 2006.

\bibitem[Airoldi et~al.(2007)Airoldi, Blei, Fienberg, and
  Xing]{articleairoldi2007}
E.~Airoldi, D.~Blei, S.~Fienberg, and E.~Xing.
\newblock Mixed membership analysis of high-throughput interaction studies:
  relational data.
\newblock \emph{ArXiv e-prints}, 2007.

\bibitem[Airoldi et~al.(2008)Airoldi, Blei, Fienberg, and
  Xing]{articleairoldi2008}
E.M. Airoldi, D.M. Blei, S.E. Fienberg, and E.P. Xing.
\newblock Mixed membership stochastic blockmodels.
\newblock \emph{Journal of Machine Learning Research}, 9:\penalty0 1981--2014,
  2008.

\bibitem[Albert and Barab\'{a}si(2002)]{articlealbert2002}
R.~Albert and A.L. Barab\'{a}si.
\newblock Statistical mechanics of complex networks.
\newblock \emph{Modern Physics}, 74:\penalty0 47--97, 2002.

\bibitem[Ball et~al.(2011)Ball, Karrer, and Newman]{articleball2011}
B.~Ball, B.~Karrer, and M.E.J. Newman.
\newblock An efficient and principled method for detecting communities in
  networks.
\newblock \emph{Phys. Rev. E}, 84\penalty0 (036103), 2011.

\bibitem[Barab\'{a}si and Oltvai(2004)]{articlebarabasi2004}
A.L. Barab\'{a}si and Z.N. Oltvai.
\newblock Network biology: understanding the cell's functional organization.
\newblock \emph{Nature Rev. Genet}, 5:\penalty0 101--113, 2004.

\bibitem[Beal and Ghahramani(2002)]{proceedingsbeal03}
M.J. Beal and Z.~Ghahramani.
\newblock The variational bayesian em algorithm for incomplete data: with
  application to scoring graphical model structures.
\newblock In JM~Bernardo, MJ~Bayarri, JO~Berger, AP~Dawid, D~Heckerman, AFM
  Smith, and M~(eds) West, editors, \emph{Bayesian Statistics 7: Proceedings of
  the 7th Valencia International Meeting}, page 453, 2002.

\bibitem[Bickel and Chen(2009)]{proceedingsbickel2009}
P.J. Bickel and A.~Chen.
\newblock A non parametric view of network models and newman-girvan and other
  modularities.
\newblock In \emph{Proceedings of the National Academy of Sciences}, volume
  106, pages 21068--21073, 2009.

\bibitem[Biernacki et~al.(2010)Biernacki, Celeux, and
  Govaert]{articlebiernacki2010}
C.~Biernacki, G.~Celeux, and G.~Govaert.
\newblock Exact and monte carlo calculations of integrated likelihoods for the
  latent class model.
\newblock \emph{Journal of Statistical Planning and Inference}, 140:\penalty0
  2991--3002, 2010.

\bibitem[Bishop(2006)]{bookbishop2006}
C.M. Bishop.
\newblock \emph{Pattern recognition and machine learning}.
\newblock Springer-Verlag, 2006.

\bibitem[Bishop and Svens\'{e}n(2003)]{proceedingsbishop2003}
C.M. Bishop and M.~Svens\'{e}n.
\newblock Bayesian hierarchical mixtures of experts.
\newblock In \emph{Proceedings of the 19th Conference on Uncertainty in
  Artificial Intelligence}, pages 57--64. U. Kjaerulff and C. Meek, 2003.

\bibitem[Blei et~al.(2003)Blei, Ng, and Jordan]{articleblei2003}
D.~Blei, A.Y. Ng, and M.I. Jordan.
\newblock Latent dirichlet allocation.
\newblock \emph{Journal of Machine Learning Research}, 3:\penalty0 993--1022,
  2003.

\bibitem[Boer et~al.(2006)Boer, Huisman, Snijders, Steglich, Wichers, and
  Zeggelink]{manualboer2006}
P.~Boer, M.~Huisman, T.A.B. Snijders, C.E.G. Steglich, L.H.Y Wichers, and E.P.H
  Zeggelink.
\newblock \emph{StOCNET : an open software system for the advanced statistical
  analysis of social networks}, 2006.

\bibitem[Daudin et~al.(2008)Daudin, Picard, and Robin]{articledaudin2008}
J.~Daudin, F.~Picard, and S.~Robin.
\newblock A mixture model for random graphs.
\newblock \emph{Statistics and Computing}, 18:\penalty0 1--36, 2008.

\bibitem[Dempster et~al.(1977)Dempster, Laird, and Rubin]{articledempster1977}
A.P. Dempster, N.M. Laird, and D.B. Rubin.
\newblock Maximum likelihood for incomplete data via the em algorithm.
\newblock \emph{Journal of the Royal Statistical Society}, B39:\penalty0 1--38,
  1977.

\bibitem[Estrada and Rodriguez-Velazquez(2005)]{articleestrada2005}
E.~Estrada and J.A. Rodriguez-Velazquez.
\newblock Spectral measures of bipartivity in complex networks.
\newblock \emph{Physical Review E}, 72:\penalty0 046105, 2005.

\bibitem[Fienberg and Wasserman(1981)]{articlefienberg1981}
S.E. Fienberg and S.~Wasserman.
\newblock Categorical data analysis of single sociometric relations.
\newblock \emph{Sociological Methodology}, 12:\penalty0 156--192, 1981.

\bibitem[Frank and Harary(1982)]{articlefrank1982}
O.~Frank and F.~Harary.
\newblock Cluster inference by using transitivity indices in empirical graphs.
\newblock \emph{Journal of the American Statistical Association}, 77:\penalty0
  835--840, 1982.

\bibitem[Gazal et~al.(2011)Gazal, Daudin, and Robin]{articlegazal2011}
S.~Gazal, J.-J. Daudin, and S.~Robin.
\newblock Accuracy of variational estimates for random graph mixture models.
\newblock \emph{Journal of Statistical Computation and Simulation}, 2011.

\bibitem[Girvan and Newman(2002)]{proceedingsgirvan2002}
M.~Girvan and M.E.J. Newman.
\newblock Community structure in social and biological networks.
\newblock In \emph{Proceedings of the National Academy of Sciences}, volume~99,
  pages 7821--7826, 2002.

\bibitem[Griffiths and Ghahramani(2005)]{proceedingsgriffiths2005}
T.~Griffiths and Z.~Ghahramani.
\newblock Infinite latent feature models and the indian buffet process.
\newblock In \emph{Neural Information Processing Systems}, volume~18, pages
  475--482, 2005.

\bibitem[Handcock et~al.(2007)Handcock, Raftery, and
  Tantrum]{articlehandcock2007}
M.S. Handcock, A.E. Raftery, and J.M. Tantrum.
\newblock Model-based clustering for social networks.
\newblock \emph{Journal of the Royal Statistical Society}, 170:\penalty0 1--22,
  2007.

\bibitem[Heller and Ghahramani(2007)]{proceedingsheller2007}
K.~Heller and Z.~Ghahramani.
\newblock A nonparametric bayesian approach to modeling overlapping clusters.
\newblock In \emph{In Proceedings of The 11th International Conference On AI
  And Statistics}, 2007.

\bibitem[Heller et~al.(2008)Heller, Williamson, and
  Ghahramani]{proceedingsheller2008}
K.~Heller, S.~Williamson, and Z.~Ghahramani.
\newblock Statistical models for partial membership.
\newblock In \emph{Proceedings of the 25th International Conference on Machine
  Learning (ICML)}, pages 392--399, 2008.

\bibitem[Hofman and Wiggins(2008)]{articlehofman2008}
J.M. Hofman and C.H. Wiggins.
\newblock A bayesian approach to network modularity.
\newblock \emph{Physical Review Letters}, 100:\penalty0 258701, 2008.

\bibitem[Holland et~al.(1983)Holland, Laskey, and
  Leinhardt]{articleholland1983}
P.~Holland, K.B. Laskey, and S.~Leinhardt.
\newblock Stochastic blockmodels: some first steps.
\newblock \emph{Social Networks}, 5:\penalty0 109--137, 1983.

\bibitem[Jaakkola and Jordan(2000)]{articlejaakkola2000}
T.S. Jaakkola and M.I. Jordan.
\newblock Bayesian parameter estimation via variational methods.
\newblock \emph{Statistics and Computing}, 10:\penalty0 25--37, 2000.

\bibitem[Jeffery(1999)]{articlejeffery1999}
C.J. Jeffery.
\newblock Moonlighting proteins.
\newblock \emph{Trends in Biochemical Sciences}, 24:\penalty0 8--11, 1999.

\bibitem[Krivitsky and Handcock(2009)]{manualkrivitsky2009}
P.N. Krivitsky and M.S. Handcock.
\newblock \emph{The latentnet package}, 2009.

\bibitem[Latouche et~al.(2009)Latouche, Birmel\'e, and
  Ambroise]{inbooklatouche2009}
P.~Latouche, E.~Birmel\'e, and C.~Ambroise.
\newblock \emph{Bayesian methods for graph clustering}, pages 229--239.
\newblock Springer, 2009.

\bibitem[Latouche et~al.(2011)Latouche, Birmel\'e, and
  Ambroise]{articlelatouche2011}
P.~Latouche, E~Birmel\'e, and C.~Ambroise.
\newblock Overlapping stochastic block models with application to the french
  political blogosphere.
\newblock \emph{Annals of Applied Statistics}, 5\penalty0 (1):\penalty0
  309--336, 2011.

\bibitem[Latouche et~al.(2012)Latouche, Birmel\'{e}, and
  Ambroise]{articlelatouche2012}
P.~Latouche, E.~Birmel\'{e}, and C.~Ambroise.
\newblock Variational bayes inference and complexity control for stochastic
  block models.
\newblock \emph{Statistical Modelling}, 12\penalty0 (1):\penalty0 93--115,
  2012.

\bibitem[Mariadassou et~al.(2010)Mariadassou, Robin, and
  Vacher]{articlemariadassou2010}
M.~Mariadassou, S.~Robin, and C.~Vacher.
\newblock Uncovering latent structure in valued graphs: a variational approach.
\newblock \emph{Annals of Applied Statistics}, 4\penalty0 (2), 2010.

\bibitem[McLachlan and Krishnan(1997)]{bookmclachlan1997}
G.~McLachlan and T.~Krishnan.
\newblock \emph{The EM algorithm and extensions}.
\newblock New York: John Wiley, 1997.

\bibitem[Newman(2006)]{proceedingsnewman2006}
M.~E.~J. Newman.
\newblock Modularity and community structure in networks.
\newblock In \emph{aaa}, volume 103, pages 8577--8582, 2006.

\bibitem[Nowicki and Snijders(2001)]{articlenowicki2001}
K.~Nowicki and T.A.B. Snijders.
\newblock Estimation and prediction for stochastic blockstructures.
\newblock \emph{Journal of the American Statistical Association}, 96:\penalty0
  1077--1087, 2001.

\bibitem[Palla et~al.(2005)Palla, Derenyi, Farkas, and
  Vicsek]{articlepalla2005}
G.~Palla, I.~Derenyi, I.~Farkas, and T.~Vicsek.
\newblock Uncovering the overlapping community structure of complex networks in
  nature and society.
\newblock \emph{Nature}, 435:\penalty0 814--818, 2005.

\bibitem[Palla et~al.(2006)Palla, Derenyi, Farkas, and Vicsek]{manualpalla2006}
G.~Palla, I.~Derenyi, I.~Farkas, and T.~Vicsek.
\newblock \emph{CFinder, the community cluster finding program}, 2006.

\bibitem[Palla et~al.(2007)Palla, Barab\'{a}si, and Vicsek]{articlepalla2007}
G.~Palla, A.L Barab\'{a}si, and T.~Vicsek.
\newblock Quantifying social group evolution.
\newblock \emph{Nature}, 446:\penalty0 664--667, 2007.

\bibitem[Snijders and Nowicki(1997)]{articlesnijders1997}
T.A.B. Snijders and K.~Nowicki.
\newblock Estimation and prediction for stochastic block-structures for graphs
  with latent block structure.
\newblock \emph{Journal of Classification}, 14:\penalty0 75--100, 1997.

\bibitem[Wang and Wong(1987)]{wang1987}
Y.J. Wang and G.Y. Wong.
\newblock Stochastic blockmodels for directed graphs.
\newblock \emph{Journal of the American Statistical Association}, 82:\penalty0
  8--19, 1987.

\bibitem[Yang and Lescovec(2013)]{proceedingsyang2013}
J.~Yang and J~Lescovec.
\newblock Overlapping community detection at scale: A nonnegative matrix
  factorization approach.
\newblock In \emph{ACM International Conference on Web Search and Data Mining
  (WSDM)}, 2013.

\bibitem[Zanghi et~al.(2008)Zanghi, Ambroise, and Miele]{articlezanghi2008}
H.~Zanghi, C.~Ambroise, and V.~Miele.
\newblock Fast online graph clustering via erd\"os renyi mixture.
\newblock \emph{Pattern Recognition}, 41\penalty0 (12):\penalty0 3592--3599,
  2008.

\end{thebibliography}

\end{document}